\documentclass[fleqn,usenatbib]{mnras}

\usepackage{newtxtext,newtxmath}
\usepackage{float}
\usepackage{placeins}
\usepackage{bm}
\usepackage{algorithm}
\usepackage{algpseudocode}
\usepackage{adjustbox}
\newcommand{\LeftComment}[1]{%
    \Statex \textcolor{blue}{\textbf{$\triangleright$} #1}%
}

\usepackage[T1]{fontenc}

\DeclareRobustCommand{\VAN}[3]{#2}
\let\VANthebibliography\thebibliography
\def\thebibliography{\DeclareRobustCommand{\VAN}[3]{##3}\VANthebibliography}

\usepackage{graphicx}	% Including figure files
\usepackage{amsmath}	% Advanced maths commands

\title[RAYTHEIA]{RAYTHEIA: A high-performance ray-tracing algorithm for three-dimensional direction-dependent equations in astronomical simulations}

\author[Zhengping Zhu et al.]{
Zhengping Zhu,$^{1}$\thanks{E-mail: zhuzhp@zhejianglab.com}
Thomas G. Bisbas,$^{1}$\thanks{E-mail: tbisbas@zhejianglab.com}
Xuefei Tang,$^{1}$
Brandt A.L. Gaches,$^{2}$
Tianwei Zhang,$^{1}$\newauthor
and Huaxi Chen$^{1}$
\\
$^{1}$Research Center for Computational Earth and Space Science, Zhejiang Laboratory, Hangzhou 311100, China \\ %Astronomical Computing, Zhejiang Laboratory
$^{2}$Faculty of Physics, University of Duisburg-Essen, Lotharstraße 1, 47057 Duisburg, Germany
}

\date{Accepted XXX. Received YYY; in original form ZZZ}

\pubyear{2025}

\begin{document}
\label{firstpage}
\pagerange{\pageref{firstpage}--\pageref{lastpage}}
\maketitle

\begin{abstract}
We present {\sc raytheia}, a high-performance reverse ray-tracing algorithm designed to efficiently solve three-dimensional direction-dependent equations in astronomical simulations. 
The algorithm uses a dual-grid framework in which the native simulation mesh -serving as the source grid for ray emission- and an adaptive mesh refinement (AMR) Cartesian contribution grid are constructed for efficient ray-walking and contribution accumulation. 
The core of the algorithm integrates a leaf-only linear-octree data structure to reduce memory overhead, the digital differential analyzer (DDA) traversal method to efficiently determine the ray-walking path, Morton Code indexing to fast leaf cell lookup during traversal, and the slab method to analytically compute the path length. Furthermore, {\sc raytheia} employs a hybrid (MPI/OpenMP) distributed parallel framework with a chunk-to-chunk communication strategy, achieving exceptional, near-ideal linear speed-up ratio and delivering high-end performance. We integrate {\sc raytheia} with the {\sc 3d-pdr} code to solve the complex chemistry and radiation transfer in photodissociation regions (PDRs). This allowed the modelling of three-dimensional PDR chemistry in a turbulent, star-forming cloud at an unprecedented resolution of $512^3$ grid cells. The algorithm demonstrates accuracy and convergence even at low angular resolutions. 
We further showcase the capabilities of {\sc raytheia} by producing high-resolution synthetic emission maps of key diagnostic lines of a star-forming region capturing physical effects such as [O{\sc i}]~$63\mu$m self-absorption, measuring the [C{\sc i}]-bright but CO-dark molecular gas, and deriving a CO-to-H$_2$ conversion factor in agreement with observations.
\end{abstract}


%%%%%%%%%%%%%%%%%%%%%%%%%%%%%%%%%%%%%%%%%%%%%%%%%%

%%%%%%%%%%%%%%%%% BODY OF PAPER %%%%%%%%%%%%%%%%%%

\section{Introduction}

% General context
Astrophysical environments are governed by a wide range of physical processes whose description requires solving equations that depend explicitly on direction. Among the most fundamental is the radiative transfer equation \citep[e.g.][]{1984frh}, which describes the propagation of photons through the medium and directly links the physical conditions of astrophysical systems to their observed emission: 
\begin{equation}
    \boldsymbol{n} \cdot \nabla I_\nu 
    = -(\kappa_{\nu,a} + \kappa_{\nu,s})\,\rho\,I_\nu 
    + j_\nu\,\rho 
    + \frac{\kappa_{\nu,s}\,\rho}{4\pi}
      \int_\Omega I_\nu\,\mathrm{d}\Omega,
    \label{eq:rte}
\end{equation}

\noindent where $\nu$ is the specific frequency, $I_\nu$ is the intensity, $\kappa_{\nu,a}$ and $\kappa_{\nu,s}$ are the absorption and scattering opacities, $\rho$ is the mass density, $j_\nu$ is the emission coefficient, $\boldsymbol{n}$ is the unit vector in the direction of propagation, and $\Omega$ is the solid angle. The time-derivative term is neglected under the assumption of a steady-state radiation field, which is appropriate for the ISM conditions considered here.

The above general form encompasses a wide range of physical regimes depending on which terms are retained. A number of ray-tracing algorithms have been developed for this specific purpose over the last two decades (see \citet{Wunsch24} for a comprehensive review). For instance, {\sc skirt} \citep{camps2013using,matsumoto2023self} solves the full scattering integral via Monte Carlo methods on Voronoi meshes, making it well suited to dust continuum and scattered-light problems \citep{Baes2022}. {\sc magritte} \cite{de2020magritte} solves the line radiative transfer problem on unstructured grids using a formal long-characteristics solver. Equation~\ref{eq:rte} has been also considered in advanced radiation-hydrodynamics simulations \citep[see e.g.][]{Bisbas15,Farcy2022,Narechania2025,Ma2025}.

Several of these algorithms use the {\sc HEALPix} package \citep{gorski2005healpix} as a guiding tool to solve such direction-dependent equations. {\sc HEALPix} offers an excellent and efficient way to tesselate the celestial sphere in equal areas and in different resolutions. The center of each, approximately square on a spherical surface, tessera defines the direction of the ray.
Ray-tracing can be then performed through three generic methods; the \textit{forward} ray-tracing, the \textit{characteristics-based} method, and the \textit{reverse} ray-tracing. 
The forward ray-tracing is used when modeling the propagation of photons emitted from a source (i.e. a massive star) to the surrounding interstellar medium (ISM) environment \citep[e.g.][]{Bisbas09,Wise11,Baczynsky15,Rosen17}. Rays are emitted from radiation sources to receiving (gas) elements, where the radiation is absorbed or scattered. Here, a ray-splitting criterion is frequently adopted allowing for a single ray emanating from the source to be split in several others (child-rays) ensuring high angular resolution at all times and for the whole domain. 
The characteristics-based method \citep[e.g.][]{Kim17, Frostholm18} is a more special way of performing ray-tracing which uses a set of rays that exist in the  domain \textit{a priori} instead of emanating them from specific sources. 
In contrast, in the reverse ray-tracing approach rays are emanated from the receiving elements (targets) rather than from the radiation sources \citep[e.g.][]{bisbas20123d,Bisbas15,clark2012treecol,wunsch2021tree,wadsley2024trevr2}. In practice, this method is favored when a large number of receiving elements (grid cells) requires integrations \textit{towards} these elements (i.e. heating functions) and/or \textit{outwards} from these elements (i.e. cooling functions). It is also ideal for simulations modeling a large number of radiation sources as discussed in \citet{wunsch2021tree}.

The chemical processes of photodissociation regions \citep[PDRs;][]{Tielens1985,Sternberg1995,Hollenbach1999,Wolfire2022} are prime examples of astrophysical environments where the angular dependence of radiation plays a central role, particularly suitable for the reverse ray-tracing approach. These regions form at the interfaces between ionized, atomic, and molecular gas, where far-ultraviolet (FUV) photons control the thermal balance, chemistry, and structure of the ISM. The penetration of FUV photons depends sensitively on both their energy and direction, determining the extent of heating and molecular dissociation in three-dimensional space. As a result, the morphology and emission properties of PDRs are strongly influenced by both the distribution of irradiating sources and the ISM density distribution. Since PDRs dominate the emission of key cooling lines, such as [C{\sc ii}], [C{\sc i}], CO and [O{\sc i}], which serve as tracers of ISM properties and conditions in both Galactic and extragalactic contexts \citep[e.g.][]{Kramer2004, Roellig2006, Langer2010, Langer2014, Pineda2013, Beuther2014, Madden2020, Zhao2024, Huang2024, Schneider2024, Topkaras25}, accurate modeling of their directional radiative transfer is essential for linking theory with observations.

In this paper we present the {\sc raytheia}\footnote{\url{https://github.com/itamos-ism/RAYTHEIA}} ray-tracing scheme, its accuracy, performance, and its integration to the {\sc 3d-pdr} code. The upgraded {\sc 3d-pdr}\footnote{\url{https://github.com/itamos-ism/3D-PDR}} is then applied to a high-resolution snapshot of a collapsing cloud representing a star-forming region, calculating the abundance distribution of species, the gas temperature distribution, and the velocity integrated emission of key coolants. {\sc raytheia} is fast, modular and flexible for extensions, making it appropriate for use in astrochemical and hydrodynamical codes.

This paper is organized as follows. Section~\ref{s:2} presents the description of the ray-casting and ray-walking approaches followed. Section~\ref{S:3} discusses the accuracy and performance of {\sc raytheia}. Section~\ref{S:4} presents the example application of PDR chemistry in a high-resolution star-forming region. We conclude in Section~\ref{s:5}.

\section{RAYTHEIA}
\label{s:2}

\subsection{Overview}

In the PDR and astrochemical context targeted by {\sc raytheia}, we use a simplified expression of Eq.~(\ref{eq:rte}) described below. We assume that scattering is negligible for the FUV and line frequencies explored here ($\kappa_{\nu,s} \approx 0$), reducing the equation to a pure absorption--emission problem \citep[see Sec.~3.8 of][for discussion on this assumption]{Bisbas23}. 
Dust enters only through the absorption opacity $\kappa_{\nu,a}$, folded into the standard FUV extinction cross-section via the dust-to-gas 
ratio. Gas-phase line emission is treated via the escape-probability formalism rather than by solving the full emission integral (see Section~\ref{s:2.4}).
Under these conditions, Eq.~(\ref{eq:rte}) reduces to:
\begin{equation}
    \boldsymbol{n} \cdot \nabla I_\nu 
    = -\kappa_{\nu,a}\,\rho\,I_\nu + j_\nu\,\rho,
    \label{eq:rte_simple}
\end{equation}

\noindent whose solution along a ray gives the well-known simplified attenuation law $I_\nu(\tau_\nu) = I_{\nu,0}\,e^{-\tau_\nu}$ \citep{clark2012treecol,okamoto2012argot,grond2019trevr}, where $I_{\nu,0}$ is the specific intensity of the radiation field at the boundary of the domain and $\tau_\nu$ is the optical depth accumulated along the ray path. This is the equation that \textsc{raytheia} discretises and integrates along each ray direction. The physical regime covered is therefore complementary to full-scattering Monte Carlo codes: {\sc raytheia} is optimised for the direction-dependent column-density and optical-depth integrations that govern PDR chemistry and thermal balance, rather than for dust scattering or non-LTE line transfer.

While neglecting diffuse radiation is in general a reasonable approximation in astrochemical calculations, we caution that such photon scattering effects may alter the molecular abundances and consequently the resulting line emission \citep[see e.g.][]{Flannery1980,LePetit2006,Goicoechea2007}. Early work by \citet{vDishoeck1984} showed that diffuse radiation can vary the photodissociation lifetime of OH by a factor of $\lesssim 2$ between isotropic and forward-scattering models. Due to the lack of dedicated studies investigating the impact of diffuse radiation on the astrochemistry of PDRs, particularly in fully three-dimensional configurations, we cannot robustly quantify how different the resulting chemical abundances and line intensities would be. Although including scattering would improve the accuracy of the radiative transfer treatment, the presented results are not expected to qualitatively change.

{\sc raytheia} addresses key limitations of earlier approaches. In particular, compared to the ray-tracing method followed in the {\sc 3d-pdr} code presented in \citet{bisbas20123d}, {\sc raytheia} achieves a substantial improvement in computational speed and memory management, enabling simulations at significantly higher spatial resolution. {\sc HEALPix} rays are emanated from every grid cells consisting the cloud to the entire domain. A key advantage of the {\sc HEALPix} scheme is its predictable gridding pattern, which ensures that every ray traces an identical solid angle. We note, however, that {\sc raytheia} can achieve equivalent efficiency with alternative user-defined methods for generating ray directions, such as a Monte Carlo approach.

{\sc raytheia} is based on the reverse ray-tracing approach. It uses a dual-grid framework (see \S\ref{S:2.1}) and considers each of the ${\cal N}_{\rm rays}=12\times4^{\ell}$ {\sc HEALPix} rays (where $\ell$ is the level of refinement) as a pencil beam along which direction-dependent equations are solved \citep[c.f.][who account for all cells overlapping the given angular cone]{clark2012treecol,wunsch2021tree,wadsley2024trevr2}. A parametric octree traversal method \citep{amanatides1987fast,revelles2000efficient,frisken2002simple} is employed 
to efficiently determine the ray-walking path, identifying the grid cells intersected by a given ray (see \S\ref{S:octree}). Ray-walk is then performed along these grid cells, which compute the integral along the ray (see \S\ref{S:intersection}). The latter is done by adopting the slab method \citep{kay1986ray, shirley2021ray, Friederichs2025}, which quickly calculates the theoretical path length of the ray through the grid cell required for integration. In addition, {\sc raytheia} uses a hybrid (MPI/OpenMP) distributed parallel framework based on domain decomposition and a chunk-to-chunk communication strategy, enabling efficient evaluation of direction-dependent integrals with excellent scalability.

\subsection{Dual-grid Parallel framework of RAYTHEIA} \label{S:2.1}
\begin{figure*}
    \centering
    \includegraphics[width=1.0\textwidth]{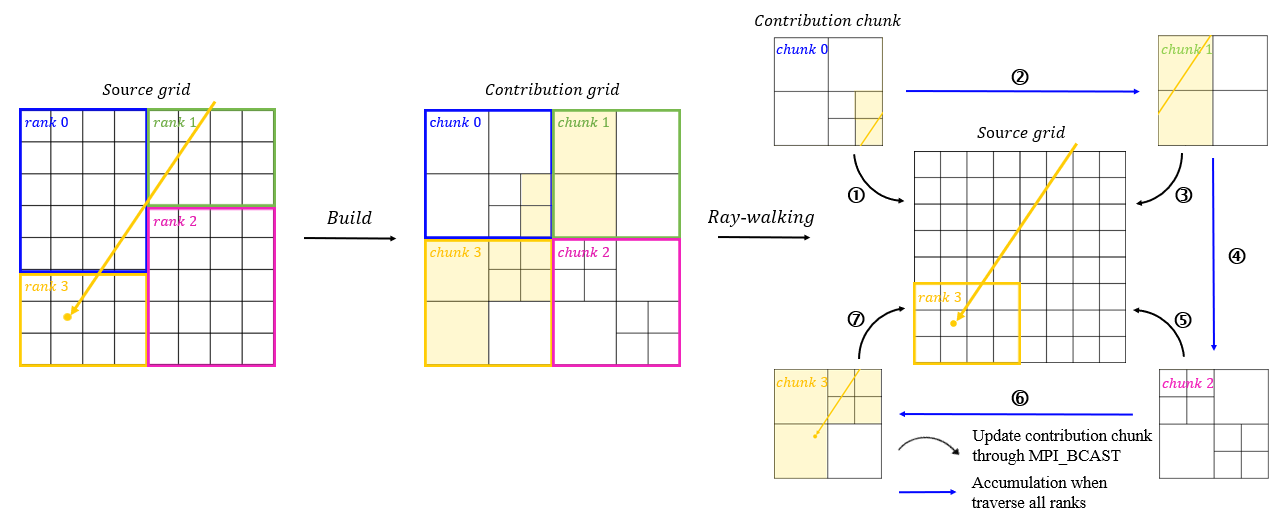}
    \caption{Two-dimensional schematic of the dual-grid parallel framework in {\sc raytheia}. Left panel: the \textit{source grid}, which can retain the native discretization of the underlying simulation, defines the locations of ray-emitting elements. Middle panel: the \textit{contribution grid}, an AMR Cartesian grid built from the source grid, provides a traversal-friendly representation for ray-walking and contribution accumulation. Right panel: calculation proceeds of the direction-dependent integral for the yellow ray emitted from rank 3. The integral is computed using a chunk-to-chunk strategy, in which contribution chunks are sequentially broadcast (black arrows), traversed to evaluate the contributions from intersected cells (yellow shading), and the resulting partial integral is accumulated into the total integral, while the same buffer is reused to receive subsequent chunks (blue arrows). The dual-grid framework is parallelized via distributed-memory domain decomposition. The computational domain is partitioned into distinct spatial chunks, each assigned to a different MPI rank (colored rectangular).}
    \label{fig:frame}
\end{figure*}

{\sc raytheia} evaluates direction-dependent integrals using a dual-grid framework composed of two decoupled grid meshes: the \textit{source grid} and the \textit{contribution grid}, as depicted in Fig.~\ref{fig:frame}. The source grid specifies the locations that rays are emitted from, while the \textit{contribution grid} is used for ray-walking and contribution accumulation.

The decoupled architecture treats the \textit{source grid} and the \textit{contribution grid} independently. The \textit{source grid} can retain the native discretization of the underlying simulation, such as structured grids, unstructured grids, or even meshless Lagrangian particles \citep{fryxell2000flash,springel2010pur,hopkins2015new}. A uniform source grid is shown as an example in Fig.~\ref{fig:frame}. The \textit{contribution grid} employs an Adaptive Mesh Refinement (AMR) Cartesian grid to optimize computational cost by confining high spatial resolution to essential regions \citep{berger1989local}. Its hierarchical, axis-aligned structure ensures highly efficient ray traversal by simplifying the path length calculation for ray-cell intersections and allowing rays to take large, adaptive steps across coarse cells \citep{amanatides1987fast,revelles2000efficient,frisken2002simple}. The dual-grid framework provides two main benefits. First, the \textit{contribution grid} provides a traversal-friendly data structure that enables efficient ray-walking. Second, the decoupled architecture allows {\sc raytheia} to accommodate different types of input data without altering the core algorithm, and support future extension of the \textit{contribution grid} to unstructured grid such as Voronoi or Delaunay meshes \citep{camps2013using,de2020magritte,lu2023extension}. The framework, thus, improves both computational efficiency and methodological generality.

The \textit{contribution grid} is built from the \textit{source grid} by recursive refinement using problem-specific refinement criteria. Once the \textit{contribution grid} has been built, the computation of the direction-dependent integral proceeds as follows. For each source element, rays are emitted from its center along a given direction. Along this ray, the algorithm performs ray-walking through the \textit{contribution grid} and searches for intersected contribution cells, as shown by the yellow shading in Fig.~\ref{fig:frame}. In each intersected cell, the local contribution to the integral is computed from the product of the cell quantity and the corresponding path length. The total integral is then obtained by accumulating the contributions from all intersected cells.

The dual-grid framework is parallelized via distributed-memory domain decomposition. The computational domain is partitioned into distinct spatial chunks, each assigned to a different MPI rank, as highlighted in different colours in Fig.~\ref{fig:frame}. Each MPI rank stores only the data associated with its local subdomain, rather than the whole dataset. Owing to the decoupled architecture, the \textit{source grid} and the \textit{contribution grid} can be decomposed differently. The \textit{source grid} can be partitioned according to the distribution of ray-emitting elements, while the \textit{contribution grid} can be decomposed to better accommodate ray traversal. This independent decomposition improves load balance.

In parallel execution, we use a chunk-to-chunk strategy, in which the direction-dependent integral is decomposed into chunk-wise partial contributions and processed one contribution chunk at a time. Only one contribution chunk is communicated and stored at each stage of the calculation, and the same buffer is reused to receive data from subsequent contribution chunks. This design reduces memory requirements and improves parallel performance by providing a regular, coarse-grained communication pattern. For a given ray, such as the yellow ray emitted from rank 3 in Fig.~\ref{fig:frame}, the direction-dependent integral calculation proceeds as follows:
\begin{enumerate} 
    \item The chunk-to-chunk strategy performs a sequential broadcast, sending each contribution chunk to rank 3 (black arrows).
    \item Once data has been received, rank 3 uses it to compute the integral contribution within its chunk. This contribution is calculated for intersected cells (yellow shadows), as described above.
    \item The same buffer is reused to receive data from subsequent contribution chunks. Consequently, the data within each contribution chunk is updated (blue arrows) after its local contribution is accumulated.
\end{enumerate}
By traversing all contribution chunks sequentially and accumulating the chunk-wise contributions, {\sc raytheia} obtains the complete direction-dependent integral. The above procedure is described in Algorithm~\ref{code} which outlines how a direction-dependent integral $\cal I$ is calculated in each MPI rank. The algorithm begins by partitioning the \textit{contribution grid} into ${\cal N}_{\rm ranks}$ contribution chunks, corresponding to the number of MPI ranks (first do-loop). Then, data corresponding to the spatial boundaries (minimum and maximum coordinates) and the integration variables of contribution chunk $c$ are broadcast to all MPI ranks. For the source elements ${\cal N}_{\rm s}$ within each MPI rank that lie inside the molecular cloud, ${\cal N}_{\rm rays}$ rays are emanated (second do-loop). For each ray, i) a parametric traversal method efficiently determines the ray-walking path, and ii) a slab method analytically computes the path length. The above process repeats until all contribution chunks have been traversed. Finally, the integral $\cal I$ is obtained by accumulating the partial contributions from all contribution chunks.

\begin{algorithm}
\caption{Pseudocode for the process of calculating the direction-dependent integration $\cal I$ in each MPI rank, where $f_k$ is the quantity to be integrated of the contribution grid cell $k$, and $r_{{\rm cell},k}$ is the path length of the ray through that cell.} \label{code}
\begin{algorithmic}[1]
\For{$c = 0$ to ${\cal N}_{\rm ranks-1}$}
\LeftComment{\textcolor{blue}{Traverse all contribution chunks}}
    \State Call \texttt{MPI\_BCAST}
\LeftComment{\textcolor{blue}{Broadcast the AMR data of contribution chunk}}
    \For{$p = 1$ to ${\cal N}_{\rm s}$}
\LeftComment{\textcolor{blue}{Traverse all source elements in source chunk}}
        \If{cell $p$ is inside the molecular cloud} 
            \For{$q = 1$ to ${\cal N}_{\rm rays}$}
\LeftComment{\textcolor{blue}{Emit rays from the cell center}}
                \State \textcolor{blue}{Determine ray-walking path within the \Statex \hspace{5.8em} contribution chunk}
                \State \textcolor{blue}{Compute path length of the ray}
                \State ${\cal I}(p,q) = {\cal I}(p,q) + f_kdr_{{\rm cell},k}$
\LeftComment{\textcolor{blue}{Accumulating the partial contribution of each contribution chunk when traversing them}}
            \EndFor
        \Else
            \State ${\cal I}(p,q) = 0$        
        \EndIf
    \EndFor
\EndFor
\end{algorithmic}
\end{algorithm}

\subsection{Computing direction-dependent integrals} \label{S:2.2}
{\sc raytheia} evaluates the direction-dependent integral by ray-walking through the AMR contribution grid and accumulating the contributions of all intersected grid cells. For a source element $p$ and ray direction $\bm q$, the discrete integral is written as
\begin{equation}
{\cal I}(p,\bm q)=\sum_k f_k\, r_{{\rm cell},k},
\end{equation}
\noindent where $f_k$ is the integrated quantity and the sum is taken over all intersected contribution grid cells, $k$.

\subsubsection{Ray-walking} \label{S:octree}
To enable efficient ray-walking, the AMR contribution grid is stored in {\sc raytheia} using a linear-octree data structure \citep{gargantini1982linear,frisken2002simple}. The octree is constructed over the spatial domain that encloses the contribution chunk, according to problem-specific refinement criteria. Cells requiring further refinement are split into eight children recursively until either the prescribed maximum octree depth is reached or the cell size decreases to the specified minimum value. The resulting terminal cells are treated as leaf cells used for ray-walking.

To support efficient indexing of leaf cells, {\sc raytheia} uses Morton Codes to map the three-dimensional logical position of each leaf cell onto a one-dimensional key while preserving spatial locality \citep{karras2012maximizing}. The linear-octree stores only the leaf cells in compact arrays containing the Morton Code, refinement level, and cell value. In this way, the search for an intersected AMR cell is converted into an array-based lookup problem rather than a recursive traversal of a pointer-based tree.

After confirming that the ray intersects the contribution chunk, {\sc raytheia} performs ray-walking on the AMR grid using the DDA method \citep{amanatides1987fast}. At each step, the advanced ray position is encoded into a Morton Code to locate the corresponding leaf cell through array lookup. The spatial extent of that cell is then reconstructed from its Morton Code and refinement level, and the ray is advanced to the next cell boundary. This procedure is repeated until the ray exits the contribution chunk. Because the step length is determined by the size of the intersected AMR cell, the traversal naturally takes large steps in coarse regions and small steps in refined regions.

The above strategy has three main advantages. First, the linear-octree stores only leaf cells, thereby reducing memory overhead compared with a fully refined grid or a pointer-based octree. Second, the combination of Morton Code indexing and the DDA method restricts the traversal to the AMR cells actually intersected by the ray rather than the full grid, thereby improving ray-walking efficiency. Third, the traversal automatically takes large steps in coarse regions and small steps in refined regions, which further improves ray-walking efficiency. This strategy is particularly advantageous in astrophysical applications with large low-density regions and localized dense structures, where a uniform-grid traversal would waste substantial effort in empty or weakly varying parts of the domain. 

\subsubsection{Calculation of path length}
\label{S:intersection}
\begin{figure}
    \centering
    \includegraphics[width=0.45\textwidth]{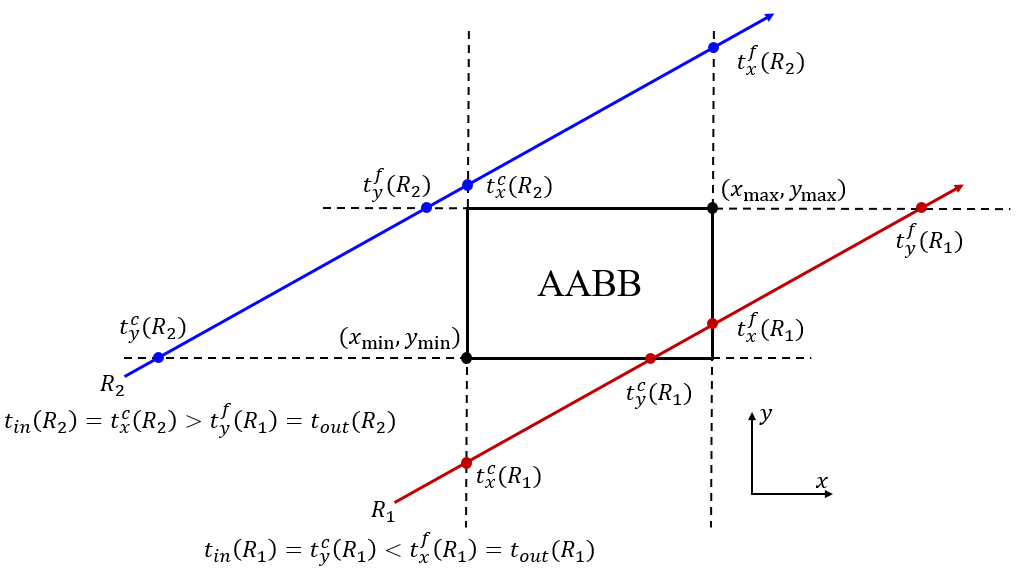}
    \caption{Two-dimensional schematic of the slab method. The axis-aligned bounding box (AABB) is defined by corners $(x_{\rm min},y_{\rm min})$ and $(x_{\rm max},y_{\rm max})$. Two rays are denoted by $R_1$ and $R_2$, respectively, and the intersection points with the slab are denoted by $t$. $R_1$ intersects with AABB rather than $R_2$ since the criterion for the ray-AABB intersection is $t_{in}<t_{out}$, where $t_{in}={\rm max}(t_x^c,t_y^c,t_z^c)$ and $t_{out}={\rm min}(t_x^f,t_y^f,t_z^f)$.}
    \label{fig:slab}
\end{figure}

Once the sequence of intersected AMR cells has been determined by the ray-walking procedure, the exact path length within each cell needs to be calculated. Geometrically, this problem reduces to computing the intersection between a ray and an axis-aligned bounding box (AABB). {\sc raytheia} utilizes the slab method \citep{kay1986ray,shirley2021ray,Friederichs2025}, which provides an efficient and analytically exact treatment of ray-box intersections. Figure~\ref{fig:slab} illustrates the slab method in two dimensions. A slab is defined as the spatial region bounded by two parallel planes,
\begin{equation}
P_i \cdot \hat{\bm n}_i  =d_i,
\label{eq:plane}
\end{equation}

\noindent where $P_i$ is any point on the plane $i$, $\hat{\bm n}_i$ is the normal vector of the plane $i$, and $d_i$ is the distance from the origin to the plane $i$. AABBs can be viewed as the volume enclosed by three distinct slabs. A ray starting from a point $Q$ can be expressed as
\begin{equation}
R = \hat{\bm m}t + Q,
\label{eq:ray}
\end{equation}

\noindent where $\hat{\bm m}$ is the direction vector of the ray. If the ray intersects the plane $i$, $P_i = R$, and the solution $t_i$ is computed by substituting equation \eqref{eq:ray} into equation \eqref{eq:plane} which yields
\begin{equation}
t_i = \frac{d_i-\hat{\bm n}_i \cdot Q}{\hat{\bm n}_i \cdot \hat{\bm m}}.
\end{equation}

\noindent For an AABB represented by two vertices $(x_{\rm min},y_{\rm min},z_{\rm min})$ and $(x_{\rm max},y_{\rm max},z_{\rm max})$ on the diagonal, there will be six intersections:
\begin{align}
&t_x^{c} = \frac{x_{\rm min}-Q_x}{m_x}, t_y^{c} = \frac{y_{\rm min}-Q_y}{m_y}, t_z^{c} = \frac{z_{\rm min}-Q_z}{m_z} \\
&t_x^{f} = \frac{x_{\rm max}-Q_x}{m_x}, t_y^{f} = \frac{y_{\rm max}-Q_y}{m_y}, t_z^{f} = \frac{z_{\rm max}-Q_z}{m_z}.
\end{align}

\noindent where the superscripts $c$ and $f$ of $t$ indicate intersections closer and farther from the origin, respectively, the subscripts $x$, $y$, and $z$ of $t$ denote the normal vector of the plane where the intersection lies, $(Q_x, Q_y, Q_z)$ is the coordinate of $Q$, $(m_x,m_y,m_z)$ are the three components of the vector $\hat{\bm m}$. The slab method cuts the ray into segments using each pair of slabs. If any segment of the ray remains in the spatial region of the AABB, it intersects the AABB. Therefore, the criterion for the ray-AABB intersection is
\begin{equation}
t_{in}<t_{out},
\end{equation}
where $t_{in}={\rm max}(t_x^c,t_y^c,t_z^c)$ and $t_{out}={\rm min}(t_x^f,t_y^f,t_z^f)$. It ensures that the ray enters the AABB before leaving it. The entry and exit points, $R_{in}$ and $R_{out}$, of the AABB can be obtained by substituting $t_{in}$ and $t_{out}$ into \eqref{eq:ray}, receptively. Subsequently, the path length $r_{cell}$ can be theoretically determined using the Euclidean norm:
\begin{equation}
r_{cell}=\left \| R_{out}-R_{in} \right \| _2=t_{out}-t_{in}.
\end{equation}

\subsection{Integrated {\sc raytheia} with {\sc 3d-pdr}}
\label{s:2.4}

\begin{figure*}
    \centering
    \includegraphics[width=0.8\textwidth]{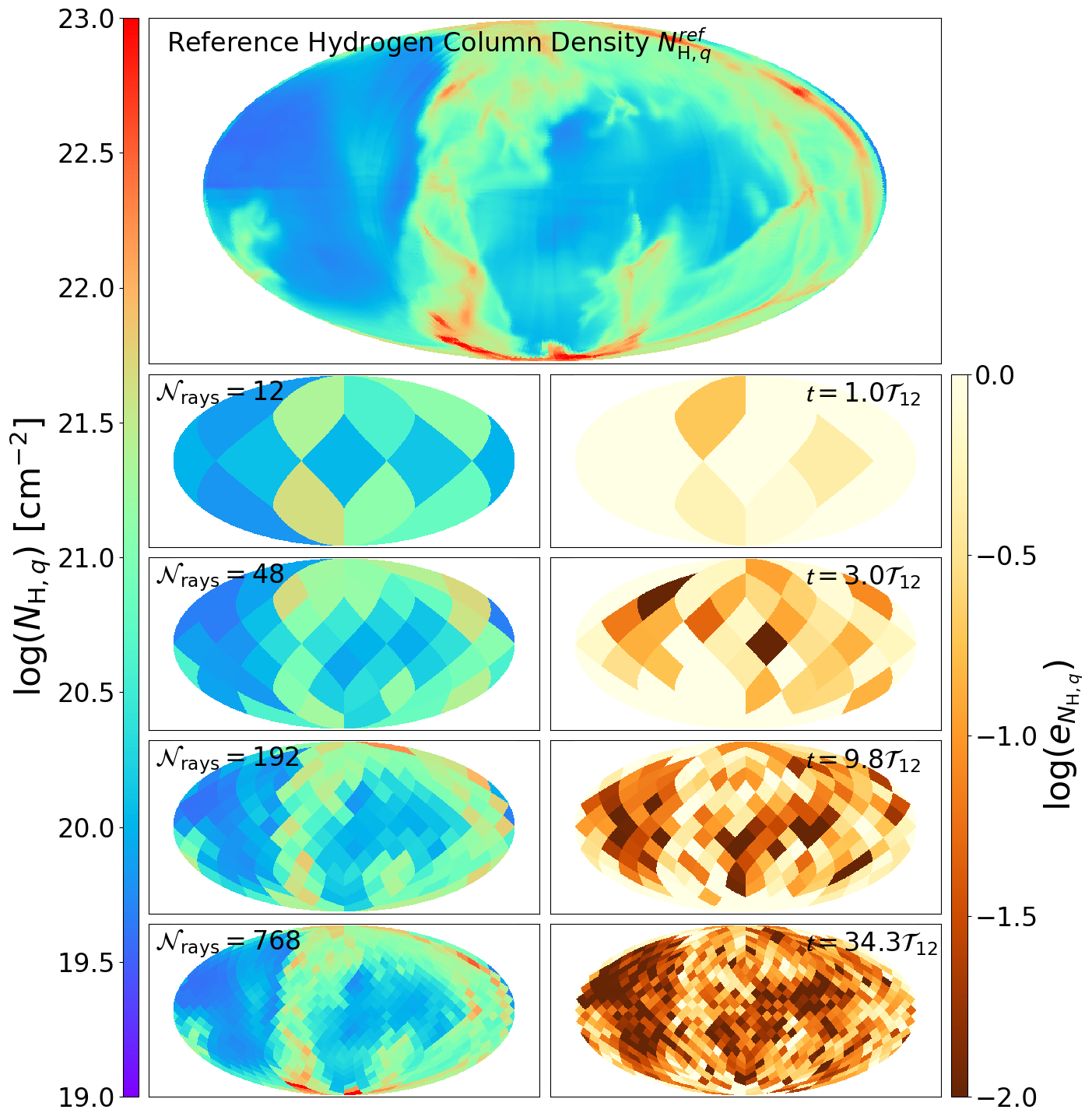}
    \caption{Accuracy of the hydrogen column density $N_{\mathrm{H},q}$ as seen from the centre of the computational domain with different angular resolution. Top panel shows the reference hydrogen column density \smash{$N_{\mathrm{H},q}^{ref}$} displayed in the Mollweide projection. Left panels below show $N_{\mathrm{H},q}$ calculated by {\sc raytheia}. The number of rays, ${\cal N}_{\rm rays}$, is shown in upper left corners of the panels. Right panels below show the relative error in the hydrogen column density $e_{N_{\mathrm{H},q}}$ calculated using Eqn.(\ref{eq:error_cd}). The elapsed time $t$ of the calculation with a single processor is shown in upper right corners of the panels, where ${\cal T}_{12}=3.68 \times 10^{-4} \rm s$ is the elapsed time with ${\cal N}_{\rm rays}=12$.}
    \label{fig:cd_accuracy}
\end{figure*}

In the present implementation, the underlying {\sc 3d-pdr} simulation is defined on a uniform Cartesian grid, which serves as the \textit{source grid} for ray emission. {\sc raytheia} then constructs an AMR Cartesian \textit{contribution grid} from it to accelerate the direction-dependent integrations, thereby pushing the achievable resolution of {\sc 3d-pdr} to previously unreachable levels. The AMR \textit{contribution grid} is constructed using density- and uniformity-based refinement criteria. A cell is further subdivided only if its maximum hydrogen number density satisfies $n_{\rm H,max} \geq n_{\rm H,PDR}$, where $n_{\rm H,PDR}$ represents the minimum PDR density of the distribution (here set to $n_{\rm H,PDR}=1\,{\rm cm}^{-3}$), and its relative density variation satisfies $(n_{\rm H,max}-n_{\rm H,min})/n_{\rm H,max}\ge 1\%$. Otherwise, it is retained as a leaf cell. The resulting refinement places spatial resolution in dense, non-uniform regions while avoiding unnecessary refinement elsewhere.

Within the coupled framework, the resulting AMR \textit{contribution grid} is used to perform the direction-dependent integrations required by {\sc 3d-pdr}, including the evaluation of the attenuation of the radiation field, the optical depth, and the escape probability entering the chemistry and thermal balance calculations.

The attenuation of the radiation field is calculated by $I_\nu(\tau_\nu) = I_{\nu,0}\,e^{-\tau_\nu}$. For a single frequency bin, the optical depth, $\tau$, and escape probability, $\beta$ \citep[describing the probability that a photon carrying a particular energy will escape the cloud without interacting with it contributing in its total cooling][]{Sobolev1960,Castor1970,deJong1975}, are calculated using the following equations \citep[e.g.][]{bisbas20123d}:
\begin{equation}
\tau (p,\bm q, i, j)=\frac{A_{ij}c^3}{8\pi\nu _{ij}^3} \frac{1}{\Delta u(p)}\int_{0}^{L} \left [ \frac{n_jg_i}{g_j}-n_i  \right ]dr,
\label{eq:tao}
\end{equation}

\begin{equation}
\beta(p) = \frac{1}{4\pi} \int_{0}^{4\pi} \frac{1 - e^{-\tau_\nu}}{\tau_\nu}d\Omega,
\label{eq:beta}
\end{equation}
where the integration is performed along the ray emanating from the source $p$ in direction $\bm q$ up to a distance $L$,
%The term $A_{ij}c^3/8\pi\nu _{ij}^3$ before the integral is a constant, 
$n_i$ and $n_j$ are the level populations of levels $i$ and $j$, respectively, and $g_i$ and $g_j$ their statistical weights. The term $\Delta u(p)=(8k_{\rm B}T(p)/\pi m_{\rm H}+v_{\rm turb}^2)^{1/2}$ is the root-mean-square of the thermal and turbulent velocities, where $k_b$ is the Boltzmann constant, $m_h$ is the proton mass, $T(p)$ is the gas temperature, and $v_{\rm turb}$ is the user-defined turbulent velocity. 

Within {\sc raytheia}, Eqns~\ref{eq:tao} and \ref{eq:beta} are discretized as follows:
\begin{equation}
\tau (p,\bm q, i,j)=\frac{A_{ij}c^3}{8\pi\nu _{ij}^3} \frac{1}{\Delta u(p)} \sum_k \left [ \frac{n_jg_i}{g_j}-n_i  \right ]_k r_{{\rm cell},k},
\label{eq:d_tao}
\end{equation}

\begin{equation}
\beta(p) = \frac{1}{{\cal N}_{\rm rays}} \sum_{q=0}^{{\cal N}_{\rm rays}} \left [ \frac{1 - e^{-\tau(q)}}{\tau(q)} \right ], 
\end{equation}
where the sum is taken over all intersected contribution grid cells $k$ and $r_{{\rm cell},k}$ is the path length of the ray through that cell. The escape probability is then computed on a direction-by-direction basis along the {\sc HEALPix} rays using the above two expressions. This formulation preserves the intrinsic anisotropy of the 3D geometry and differs from approximations which potentially assume a specific symmetry (e.g. expanding sphere).

\section{Accuracy and performance of {\sc raytheia}}
\label{S:3}

To explore the accuracy and performance of {\sc raytheia}, we construct a test\footnote{All tests described in \S\ref{S:3} and \S\ref{S:4} were performed on the Sugon HPC cluster at Zhejiang Lab. The cluster has 196 nodes, each equipped with two 28-core Intel Xeon Gold 6348 @ 2.6GHz processors and 1 Terabyte memory.} similar to the one described in \citet{clark2012treecol} and \citet{wunsch2021tree} where we calculate the sky map of the total hydrogen column density at different {\sc HEALPix} refinements (and therefore angular resolutions) as seen from the center of the computational domain from a given density distribution.

We employ a density distribution representing a collapsing molecular cloud (see \S\ref{S:4}). The original hydrodynamical simulation used a $512^3$ uniform grid, however for the purposes of the test in this section we downgraded it to $128^3$. 

The accuracy and convergence of {\sc raytheia} are examined by comparing the hydrogen column density under four angular resolutions, with ${\cal N}_{\rm rays}=12$, 48, 192, and 768 rays (corresponding to $\ell=0-3$ {\sc HEALPix} levels of refinement). Our reference result is the one obtained under very high angular resolution ${\cal N}_{\rm rays}=49152$ ($\ell=6$). The relative error in the hydrogen column density is defined as
\begin{equation}
e_{N_{\mathrm{H},q}}=\frac{\left | N_{\mathrm{H},q}- N_{\mathrm{H},q}^{ref} \right |}{N_{\mathrm{H},q}^{ref}},
\label{eq:error_cd}
\end{equation}
where $N_{\mathrm{H},q}$ is the hydrogen column density along ray $q$, \smash{$N_{\mathrm{H},q}^{ref}$} is the mean value of reference hydrogen column density along ray $q$.

Figure~\ref{fig:cd_accuracy} summarizes the results, showing Mollweide projections of \smash{$N_{\mathrm{H},q}^{ref}$} (top panel), $N_{\mathrm{H},q}$ (left panels), and $e_{N_{\mathrm{H},q}}$ (right panels) along different rays. The angular resolution (${\cal N}_{\rm rays}$) and elapsed time ($t$) of the calculation with a single processor are indicated in the upper left and upper right corner of each panel, respectively. Even at a low resolution of ${\cal N}_{\rm rays}=12$, {\sc raytheia} recovers the overall density distribution of the turbulent cloud. As the angular resolution increases, the algorithm reveals finer structural details and reduces the error $e_{N_{\mathrm{H},q}}$. This improvement in accuracy comes at a linear computational cost, as evidenced by the increase in elapsed time.

For each cell, we define the effective visual extinction $A_{\rm V,eff}$ as the local visual extinction averaged over all ray directions. Physically, $A_{\rm V,eff}$ represents the average FUV shielding experienced by the cell from the ambient radiation field, accounting for the full three-dimensional distribution of surrounding material. 
This quantity is calculated as \citep{glover2010modelling,offner2013modeling}:
\begin{equation}
A_\mathrm {V,eff}=-\frac{1}{\gamma} \ln \left ( \frac{1}{\cal N_{\rm rays}} \sum_{q=0}^{\cal N_{\rm rays}} e^{-\gamma A_{\rm V_{\rm o}} N_{\rm{H},q}} \right ),
\end{equation}
where $\gamma=3.02$ is the dust attenuation factor 
%\citep{bergin2004molecular,van2006photoprocesses,Roellig2007}, 
and $A_{\rm V_{\rm o}}=6.29\times 10^{-22}$ mag cm$^2$ \citep{Roellig2007}. The relative error is then given by:
\begin{equation}
e_{A_\mathrm {V,eff}}=\frac{\left | A_\mathrm {V,eff} - A_\mathrm {V,eff}^{ref} \right |}{A_\mathrm {V,eff}^{ref}},
\end{equation}
where $A_\mathrm {V,eff}^{ref}$ is a reference value computed using the high-resolution hydrogen column density, \smash{$N_{\mathrm{H},q}^{ref}$}. 
We note that $A_{\rm V,eff}$ is not directly used during the chemical calculations in {\sc 3d-pdr}. Instead, the reaction rates are computed independently along each individual {\sc HEALPix} ray direction and subsequently averaged. The $A_{\rm V,eff}$ quantity is therefore introduced only as a convenient direction-averaged measure of the local shielding environment and it is used for comparison with other averaged quantities.

\begin{figure}
    \centering
    \includegraphics[width=0.45\textwidth]{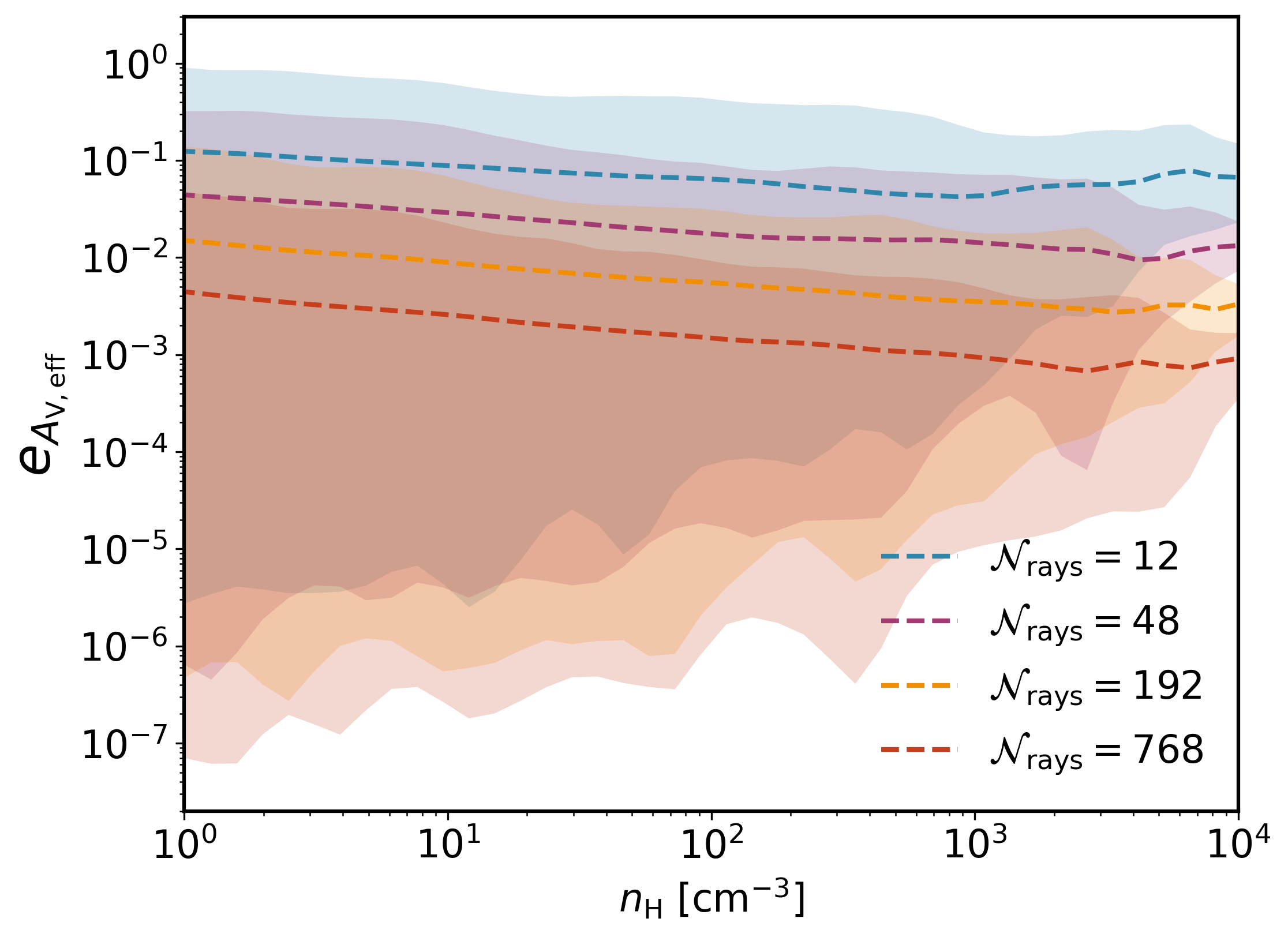}
    \caption{The $e_{A_\mathrm {V,eff}}$ profiles for different angular resolutions. The shaded area shows the range of $e_{A_\mathrm {V,eff}}$ that takes all grid cells into account. The dashed line represents the mean profile of $e_{A_\mathrm {V,eff}}$.}
    \label{fig:aveff}
\end{figure}

\begin{figure}
    \centering
    \includegraphics[width=0.45\textwidth]{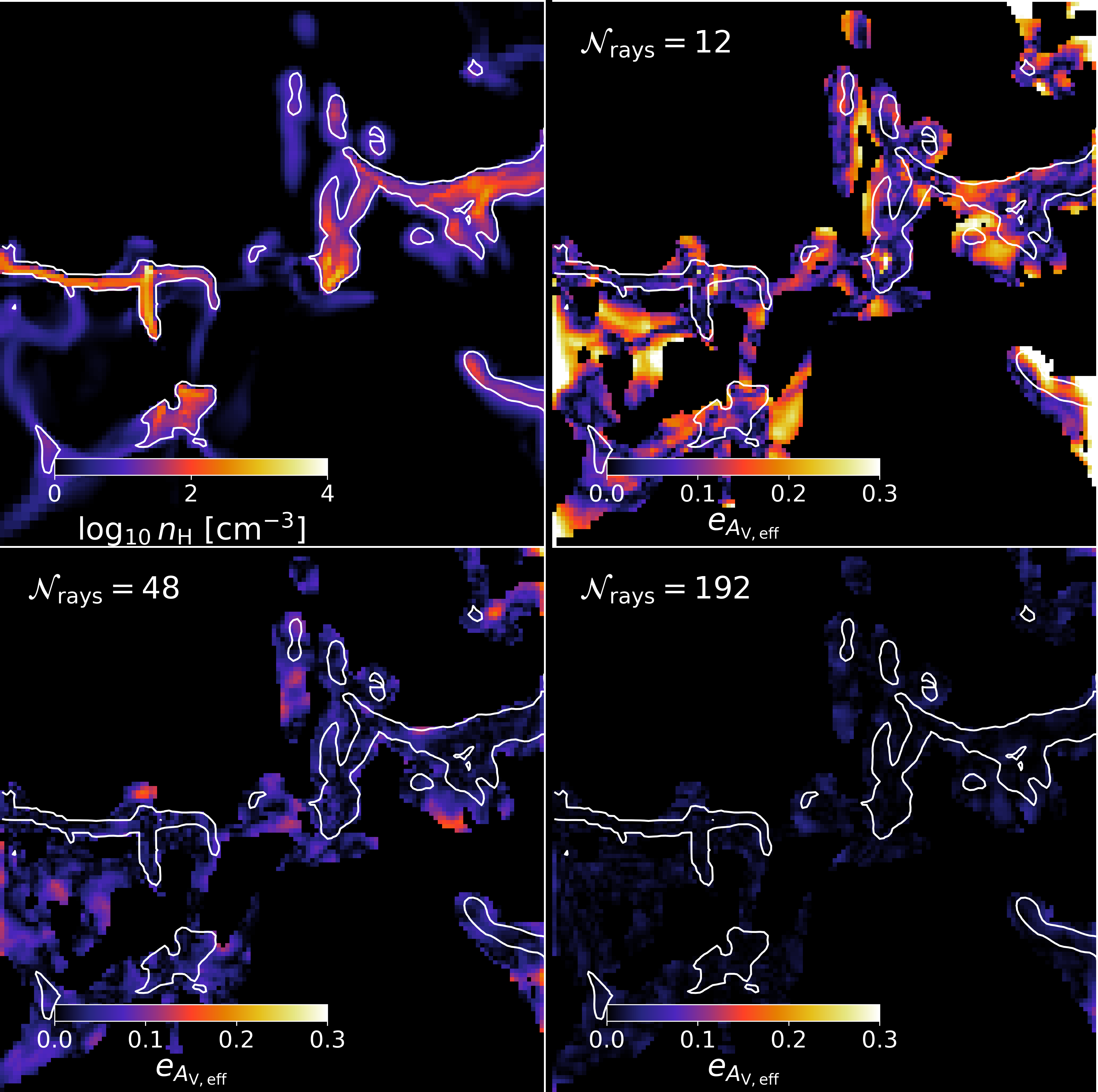}
    \caption{Spatial distribution of the $e_{A_{\rm V,eff}}$ on the $x$-$z$ plane at the domain centre. Top-left: the local number density $\log_{10} n_{\rm H}$ of the slice. The remaining panels show $e_{A_{\rm V,eff}}$ at $\mathcal{N}_{\rm rays}=12$ (top right), 48 (bottom left), and 192 (bottom right). The white contour marks $n_{\rm H} = 10$~cm$^{-3}$, corresponding to the dense gas region where PDR chemistry is most active. The spatial average of $e_{A_{\rm V,eff}}$ within this region is 0.08, 0.02, and 0.01 for $\mathcal{N}_{\rm rays}=12$, 48, and 192, respectively, demonstrating clear spatial convergence as the angular resolution increases.}
    \label{fig:av_spatial}
\end{figure}

Figure~\ref{fig:aveff} demonstrates the accuracy of {\sc raytheia} in evaluating $A_\mathrm{V,eff}$ versus the local number density, $n_{\rm H}$. For each angular resolution, the dashed line represents the mean profile of $e_{A_\mathrm{V,eff}}$, with the shaded area indicating its range across all $128^3$ grid cells. Both the mean profile and the shaded area decrease systematically with increasing $n_{\rm H}$ and $\mathcal{N}_{\rm rays}$, demonstrating good accuracy and convergence properties of \textsc{raytheia}. However, by binning all cells by density, spatial information is lost. To complement this, Fig.~\ref{fig:av_spatial} shows the spatial distribution of $e_{A_{\rm V,eff}}$ on the $x$-$z$ plane at the domain centre at $\mathcal{N}_{\rm rays}=12$, 48, and 192. The white contour marks $n_{\rm H} = 10$~cm$^{-3}$, mapping out the dense gas region where the PDR chemistry is most active and where the accuracy of $A_{\rm V,eff}$ is, therefore, most critical for the {\sc 3d-pdr}. The largest errors at $\mathcal{N}_{\rm rays}=12$ are concentrated in the diffuse regions outside the white contour, where the limited number of ray directions is less able to capture the anisotropic column density distribution. Inside the dense region, the spatial average of $e_{A_{\rm V,eff}}$ decreases from 0.08 at $\mathcal{N}_{\rm rays}=12$ to 0.02 at $\mathcal{N}_{\rm rays}=48$ and 0.01 at $\mathcal{N}_{\rm rays}=192$, and the errors remain small even at $\mathcal{N}_{\rm rays}=12$. Together, Figs.~\ref{fig:aveff} and~\ref{fig:av_spatial} demonstrate that $e_{A_{\rm V,eff}}$ converges clearly with increasing $\mathcal{N}_{\rm rays}$, both statistically and spatially.


Furthermore, we investigate how key outputs of {\sc 3d-pdr}, such as chemical abundances and gas temperatures, depend on the angular resolution. We test the {\sc 3d-pdr} response by modeling PDR chemistry at three {\sc HEALPix} refinement levels ($\ell=0,1,2$), corresponding to 12, 48, and 192 rays, respectively. As shown in Appendix~\ref{app:abundances}, the abundances of C+, C, CO, H,{\sc i}, and H$2$, along with the gas temperature, show minimal variation with ${\cal N}_{\rm rays}$ across local number densities $n_{\rm H}$. This consistency indicates excellent convergence. Given that the computational cost for ${\cal N}_{\rm rays}=12$ is a third of that for 48 rays, the minor discrepancies are well justified by the substantial gain in efficiency.

The preceding tests demonstrate that ${\cal N}_{\rm rays}=48$ provides a good compromise between accuracy and computational cost, serving as a suitable default choice for most applications. While ${\cal N}_{\rm rays}=12$ is less precise at resolving directional structures, the resulting chemical abundances and gas temperatures show only minor deviations from those at higher angular resolution. This is because the chemistry and thermal balance in {\sc 3d-pdr} depend primarily on angle-averaged quantities that are captured reasonably well at ${\cal N}_{\rm rays}=12$ \citep{bisbas20123d,wunsch2018tree,Bisbas23}. Therefore, ${\cal N}_{\rm rays}=12$ remains a cost-effective alternative for {\sc 3d-pdr} when the primary goal is to obtain converged chemical and thermal properties rather than highly resolved directional maps.

\begin{figure}
    \centering
    \includegraphics[width=0.45\textwidth]{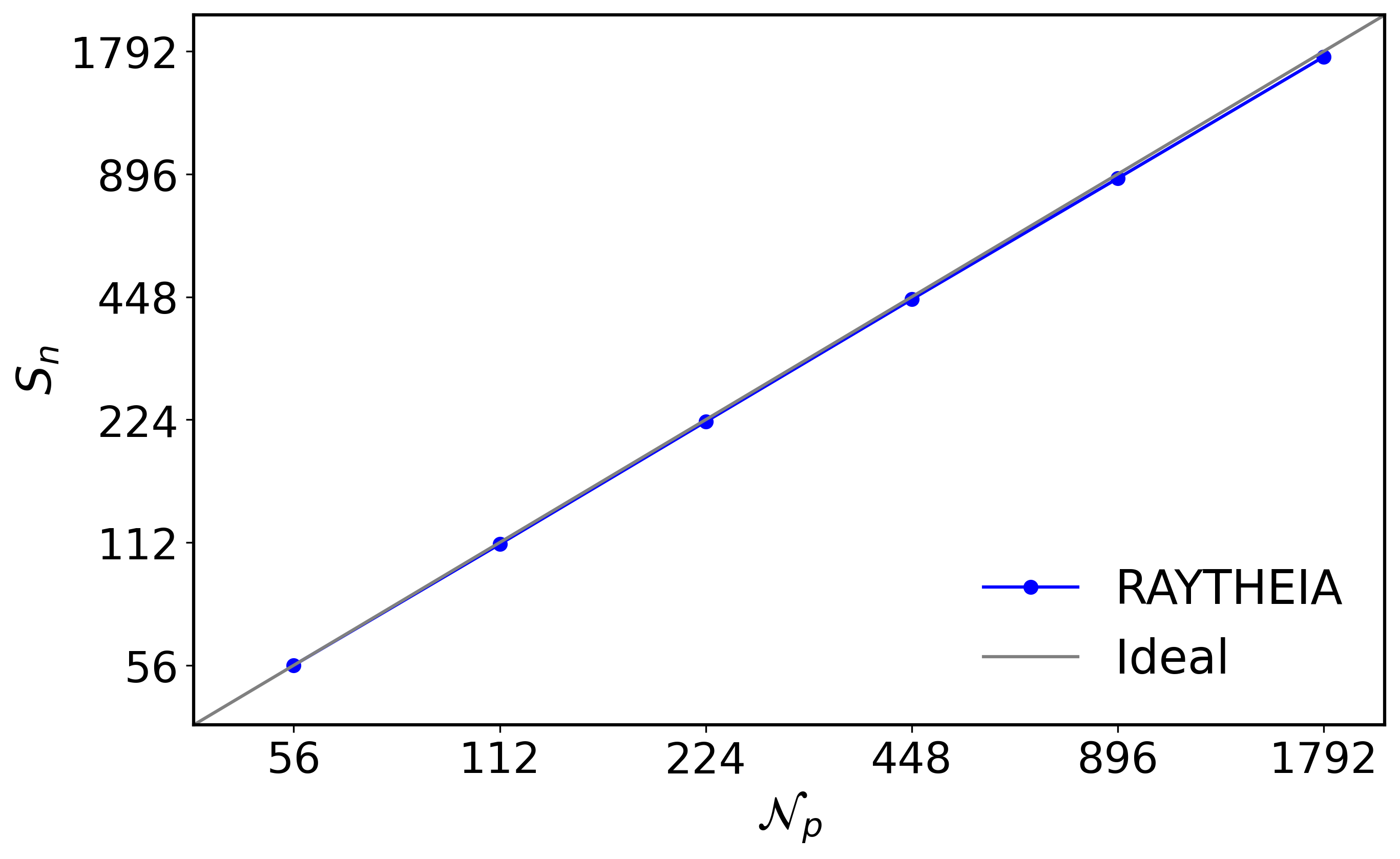}
    \caption{Strong scaling test. Speed-up ratio as a function of the total number of processors on up to 1792 for the turbulent cloud with $128^3$ grids. The grey line shows the ideal scaling $S_n = {\cal N}_p$.}
    \label{fig:scaling}
\end{figure}

We performed a strong scaling test to evaluate the performance of {\sc raytheia} in computing $A_\mathrm{V,eff}$ for all grid cells. The test utilizes an MPI/OpenMP hybrid parallelization framework, where the total number of processors is ${\cal N}_p = {\cal N}_{\rm ranks} \times {\cal N}_{\rm threads}$. Here, ${\cal N}_{\rm ranks}$ is the number of MPI ranks, and ${\cal N}_{\rm threads}=56$ is the number of OpenMP threads per MPI rank on our testing platform at Zhejiang Lab. The speed-up ratio $S_n$ on ${\cal N}_p$ processors is defined with respect to the 56-processor run as:
\begin{equation}
S_n=\frac{t_{56}}{t_p},
\end{equation}
where $t_{56}$ and $t_{\rm p}$ are the elapsed times using 56 and ${\cal N}_p$ processors, respectively. As shown in Figure~\ref{fig:scaling}, the speed-up ratio of {\sc raytheia} is almost indistinguishable from the ideal linear scaling, demonstrating its excellent performance in massively parallel computations.

\section{Example application}
\label{S:4}

To demonstrate the capabilities of {\sc raytheia}, we perform a high-resolution astrochemical model of a star-forming region using the upgraded {\sc 3d-pdr} code. The region (used also in \S\ref{S:3}) is taken from the SILCC-Zoom project \citep{Seifried17} and represents a collapsing molecular cloud with mass $M\simeq1.5\times10^5\,{\rm M}_{\odot}$. The domain size is 125~pc and the resolution of the uniform-grid is $512^3$, thus each grid cell has a size of $\sim0.24\,{\rm pc}$. The selected region is part of a zoom-in simulation of molecular cloud formation within a stratified, supernova-driven, multiphase galactic disc, where the cloud assembles self-consistently out of the diffuse ISM under the combined effects of turbulence, self-gravity, and large-scale flows. In the SILCC models, the cloud has been dynamically evolved with supersonic turbulence inherited from the galactic environment and continuously accreting material leading to a highly structured, filamentary morphology. Although the simulations of \citet{Seifried17} include several chemical and dynamical processes, we use here only the density and velocity distributions -- the latter to construct realistic synthetic maps.

In the {\sc 3d-pdr} simulation we use an $\ell=0$ HEALPix refinement level\footnote{
Although $\ell=1$ is the preferred default for most applications, we retain $\ell=0$ here to reduce the computational expense; the full coupled calculation at $512^3$ resolution already requires approximately $2\times10^5$ CPU hours and increasing to $\ell=1$ would triple this cost. The convergence tests in Appendix~A demonstrate that $\ell=0$ produces chemical abundances and gas temperatures consistent with those at $\ell=1$ and $\ell=2$, justifying this choice for the present high-resolution astrochemical calculation.} and a 33-species, 330-reaction subset of the UMIST2012 chemical network \citep{McElroy2013}. Adopting a total hydrogen density threshold of $n_{\rm H} \geq 1\,{\rm cm}^{-3}$ to flag the PDR cells, we find that $\sim25\%$ of all cells can be modeled with chemistry related to PDRs. The rest of the gas is very rarefied and ionized, thus not modeled here.

Regarding the choice of ISM environmental conditions for the PDR modeling, we assumed an isotropic radiation field of strength of $\chi/\chi_0=10$ \citep[treated as a single parameter and normalized to the spectral shape of][]{Draine78} in which the FUV radiation impinges radially. This means that the FUV radiation is treated as a uniform illumination from all directions sampled via inward-directed rays. We use solar metallicity and a dust-to-gas ratio of $10^{-2}$. The cosmic-ray ionization rate is computed using the $\mathcal{H}$-model of \citet{Padovani18}, with a normalization in the outer cloud (where $n_{\rm H}<1\,{\rm cm}^{-3}$) corresponding to $\log\zeta_{\rm CR}=-15.2$. Cosmic rays are assumed to be isotropic and their attenuation is treated directionally, in analogy to the FUV field, by computing column densities along {\sc HEALPix} rays as implemented in \citet{Gaches2022}.

\subsection{Column density and velocity integrated maps}

 \begin{figure*}
     \centering
     \includegraphics[width=0.95\linewidth]{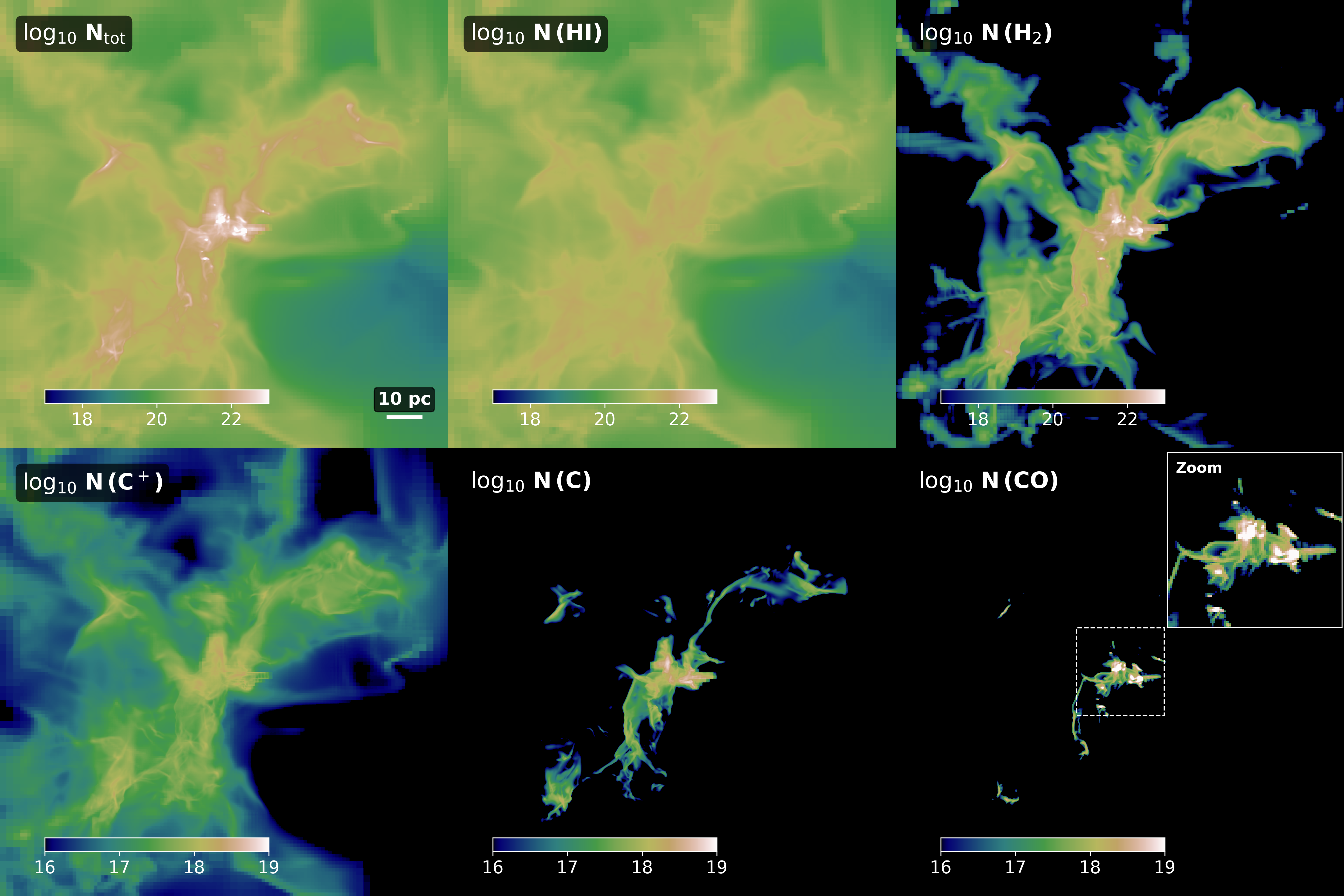}
     \caption{Column density plots of the star-forming region taken from the SILCC-Zoom project \citep{Seifried17}. Top row (left-to-right): total H-nucleus column density, H{\sc i} column density, H$_2$ column density. Bottom row: C$^+$ column density, C column density and CO column density. The inset on the bottom right panel shows a close-up of the central region in N(CO).}
     \label{fig:cds}
 \end{figure*}

Figure~\ref{fig:cds} shows column density maps of the total H-nucleus column density ($\rm N_{\rm tot}$), the atomic and molecular hydrogen (H{\sc i}, H$_2$) column densities and the column densities of the so-called `carbon cycle' consisting of ionized carbon (C$^+$), atomic carbon (C) and carbon monoxide (CO). As expected, $\rm N(H_2)$ is associated with higher column densities of $n_{\rm H}$ since it is in these areas where the FUV radiation is extinguished allowing for the formation of H$_2$ molecule on dust grains \citep{Hollenbach1979,Cazaux04}. The carbon cycle species also follow this sequence: the outer, lower-density part of the star-forming region interacting with the FUV radiation field, is rich in C$^+$. However, as the $\rm N_{\rm tot}$ column density increases, carbon is found in atomic form (C) mainly due to the photodissociation process of CO \citep{vanDishoeck88, Sternberg1995}. In places much deeper in the cloud, it is found in the form of CO. 

  \begin{figure*}
      \centering
      \includegraphics[width=0.95\linewidth]{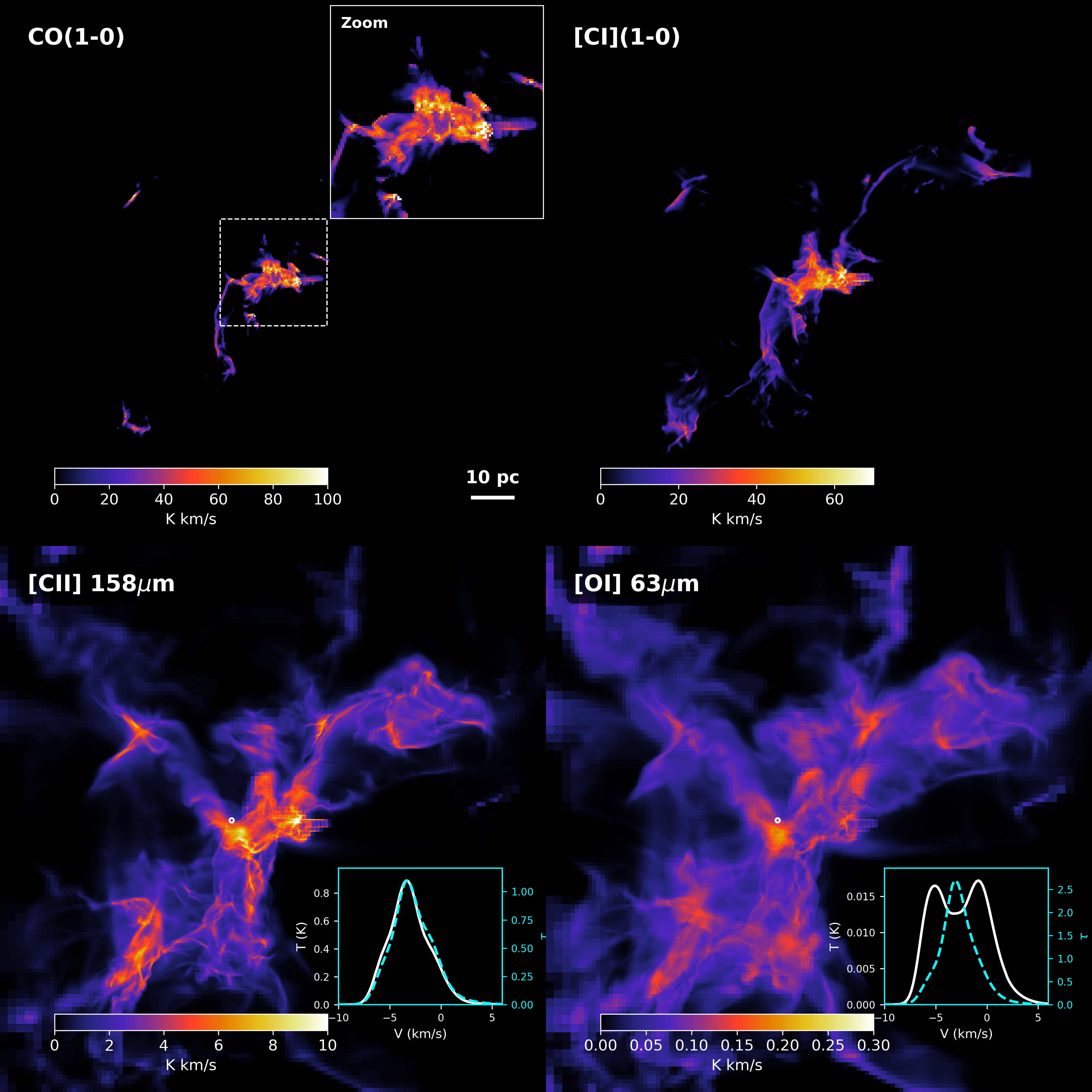}
      \caption{Velocity integrated emission of CO~(1-0) (top left), [C{\sc i}]~(1-0) (top right), [C{\sc ii}]~$158\mu$m (bottom left), and [O{\sc i}]~$63\mu$m (bottom right). The inset on the top left panel shows a close-up of the central region in CO~(1-0) emission. The insets on the bottom panels show the brightness temperature and optical depth spectra for the region marked with a white circle in the middle of the cloud.}
      \label{fig:W}
  \end{figure*}

Figure~\ref{fig:W} shows velocity integrated synthetic maps of the CO~(1-0), [C{\sc i}]~(1-0), [C{\sc ii}]~$158\mu$m, and [O{\sc i}]~$63\mu$m lines. These emission maps have been calculated by solving the radiative transfer equation along the line-of-sight of the observer by taking into account the velocity field from the hydrodynamical snapshot \citep[see][for the description of the relevant radiative transfer algorithm]{Bisbas17,Bisbas21}. As can be seen in Fig.~\ref{fig:W}, the emission of CO~(1-0) originates from the high column density region where star-formation takes place. The fine-structure line of [C{\sc i}]~(1-0) is also associated with high column densities and therefore the H$_2$-rich gas. Additionally, it follows more closely the structure of N(H$_2$) than that of CO~(1-0), as further discussed below. Both [C{\sc ii}]~$158\mu$m and [O{\sc i}]~$63\mu$m are emitted from the outermost part of the star-forming region, in agreement with observations \citep[e.g.][]{Beuther2014,Schneider2018,Mookerjea2019,Guevara2024,Okada2024}. We also find that the [O{\sc i}]~$63\mu$m emission line is considerably fainter than [C{\sc ii}]~$158\mu$m and appears more spatially extended. However, its faintest emission lies close to the detection limit, causing the line to appear more compact in observations.

\subsection{[O{\sc i}] self-absorption}

We identify self-absorption in the [O{\sc i}]~$63\mu$m line within the central, high-density star-forming region. 
Self-absorption occurs when a cold component of the cloud is along the line-of-sight and between the source of emission and the observer \citep{Liseau2006,Goldsmith2019,Goldsmith2021}.
From the small white circle (radius $=0.49\,{\rm pc}$, area $\approx0.75\,{\rm pc}^2$) in the lower panels of Fig.~\ref{fig:W}, we extract spectra of the brightness temperature ($T$, white solid line) and optical depth ($\tau$, cyan dashed line), shown in the panel insets. The velocity range is $16\,{\rm km}\,{\rm s}^{-1}$. 
While the brightness temperature of the [O{\sc i}]~$63\mu$m line exhibits two distinct peaks, its optical depth is single-peaked. This $\tau$ peak reaches a value of $\approx 2.5$ -confirming the line is optically thick- and aligns precisely with the local minimum in $T$. In contrast, the [C{\sc ii}]~$158\mu$m spectra show single peaks for both brightness temperature and optical depth, with $\tau$ peaking at $\approx1$. Given that both lines originate from gas with nearly identical physical properties \citep{Goldsmith2012,Goldsmith2019}, the double-peaked structure in [O{\sc i}] is a signature of self-absorption. This interpretation is further supported by the spectral correlation: the peak in optical depth coincides exactly with the velocity of the brightness temperature dip ($\approx -3.0$ km s$^{-1}$), a strong indicator of self-absorption in dense interstellar gas.

It would be interesting to estimate the column density of oxygen, N(O), from this self-absorption feature and compare it with the model. Following Equation 32 of \citet{Mangum2015}, we estimate N(O) to be approximately $5.2\times10^{18}\,{\rm cm}^{-2}$. This calculation assumes an excitation temperature of ${\sim}10^3\,{\rm K}$, consistent with the gas temperature in the selected region. The resulting column density is in good agreement with the value directly derived from the simulation for the same area, ${\sim}3\times10^{18}\,{\rm cm}^{-2}$. This consistency supports the presence of a self-absorption feature even at the moderate optical depths ($\tau\sim2.7$) observed here.

\subsection{The $X_{\rm CO}$ conversion factor and the `CO-dark' molecular gas}

Furthermore, as a proof-of-concept for the validity of our numerical approach, we additionally calculate the CO-to-H$_2$ conversion factor, the so-called `$X_{\rm CO}$-factor' \citep[see][for a review]{Bolatto13} defined as
\begin{eqnarray}
    X_{\rm CO}=\frac{\sum_i\rm N_i(H_2)}{\sum_i\rm W_i(CO,\,1-0)},
\end{eqnarray}
and compare it with the observed value in our Galaxy. In the above, the index $i$ runs over all source grid cells satisfying the observational limit $W_i({\rm CO,\,1-0}) \geq 0.1$~K\,km\,s$^{-1}$, following local observations \citep[e.g.][]{Leroy16,Tokuda21,Luo24}. Our model estimates a conversion factor of $X_{\rm CO}\simeq5.3\times10^{20}\,{\rm cm}^{-2}\,({\rm K}\,{\rm km}/\,{\rm s})^{-1}$, which is slightly higher than the average conversion factor observed in the Milky Way of $\simeq2\times10^{20}\,{\rm cm}^{-2}\,({\rm K}\,{\rm km}/\,{\rm s})^{-1}$ \citep{Bolatto13}.
This elevated value of $X_{\rm CO}$ found in our models does not arise from subthermal excitation of the CO(1–0) transition, as the level populations are close to LTE ($T_{\rm ex}\approx T_{\rm gas}$). Instead, it is primarily a consequence of radiative transfer effects in the high-column, collapsing gas. As described earlier, the simulated cloud spans a volume of $125^3\,\mathrm{pc}^3$, with the majority of the volume ($\sim75\%$) occupied by diffuse, non-PDR gas ($n_{\rm H}<1\,\mathrm{cm^{-3}}$), while most of the $\sim1.5\times10^5\,{\rm M_{\odot}}$ mass is concentrated within a compact, collapsing region of size $\sim15^3\,\mathrm{pc}^3$. This strong mass concentration leads to a rapid build-up of column density and, consequently, very large optical depths in CO(1–0) ($\tau_{\rm CO,1-0}\gg1$, see Appendix~\S\ref{app:tau}). In this regime, the line becomes saturated and the emergent brightness temperature is limited by the excitation temperature rather than the total column density. As a result, $\rm W(CO,1-0)$ increases only weakly with increasing H$_2$ column, while N(H$_2$) continues to grow, leading to an elevated $X_{\rm CO}$ value. Thus, CO(1–0) no longer traces the full molecular column but is primarily sensitive to the $\tau_{\rm CO,1-0}\sim1$ layers in velocity space, with additional gas effectively hidden behind optically thick material. This naturally explains the higher $X_{\rm CO}$ values obtained in dense, collapsing regions without invoking subthermal excitation or strong self-absorption effects.

Furthermore, it is known that there is a considerable fraction of H$_2$-rich ISM gas that is not bright in CO molecular lines, the so-called `CO-dark' gas \citep{vanDishoeck92}. Towards this, [C{\sc i}]~(1-0) has been proposed as an alternative tracer capable of measuring the CO-dark H$_2$-rich gas \citep{Papadopoulos04,Offner14,Bisbas21,Bisbas25}. It would be, therefore, interesting to investigate the amount of CO-dark gas component in this high-resolution model and compare it with the H$_2$ gas traced by [C{\sc i}](1-0).

\begin{figure}
    \centering
    \includegraphics[width=0.95\linewidth]{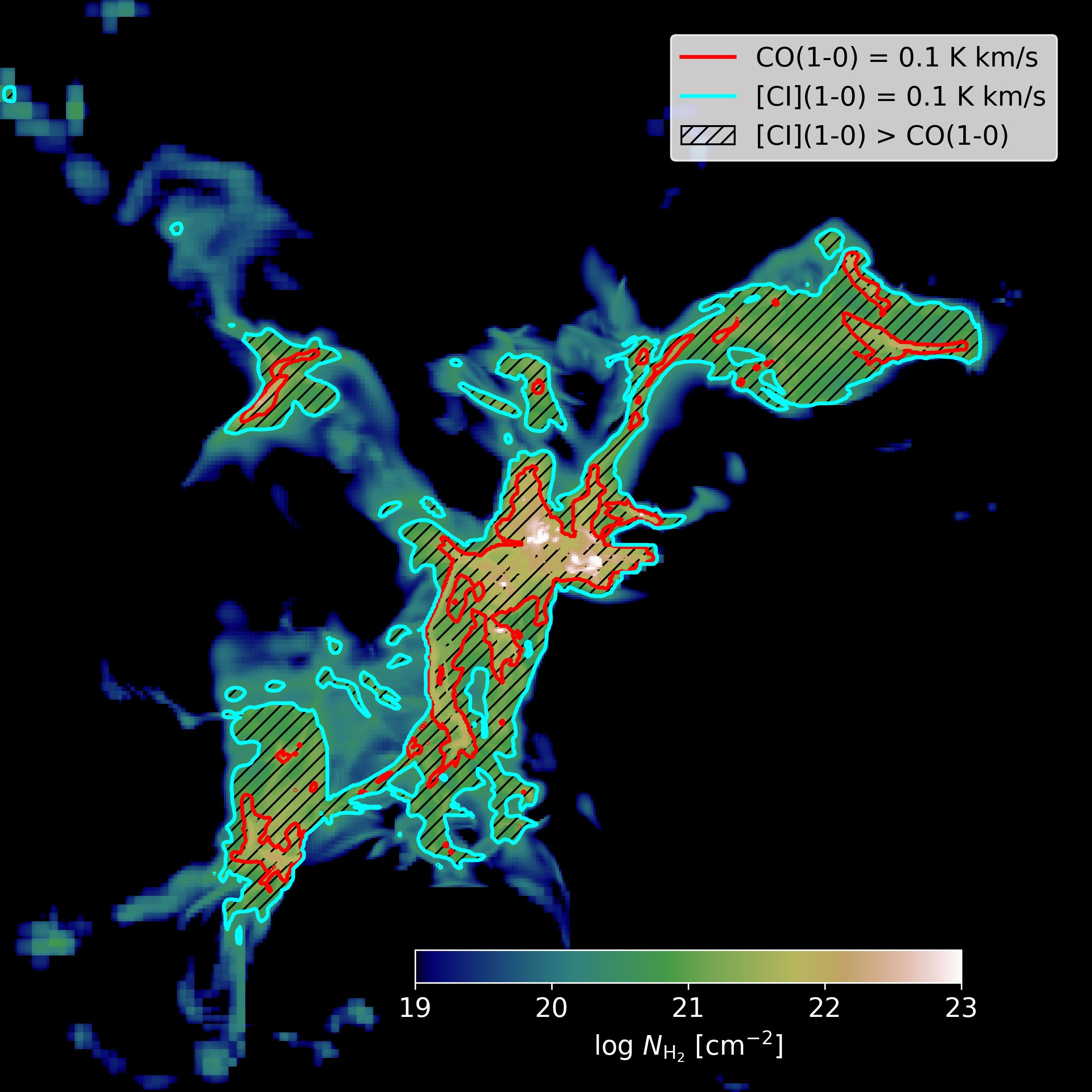}
    \caption{H$_2$ column density map with contours marking the adopted $0.1\,{\rm K}\,{\rm km}\,{\rm s}^{-1}$ detection limit for CO~(1-0) (red) and [C{\sc i}]~(1-0) (cyan). The hatched region indicates gas where the [C{\sc i}]~(1-0) intensity exceeds that of CO~(1-0). This corresponds to CO-dark gas where [C{\sc i}]~(1-0) is bright (outside the CO-bright region) and to CO-bright gas where [C{\sc i}]~(1-0) is the dominant tracer (within the CO-bright region).}
    \label{fig:codark}
\end{figure}

Figure~\ref{fig:codark} shows the N(H$_2$) map in which we mark the detection limit of $0.1\,{\rm K}\,{\rm km}\,{\rm s}^{-1}$ for [C{\sc i}]~(1-0) in cyan and for CO~(1-0) in red. The hatched region between these contours corresponds to H$_2$ gas that is CO-dark but [C{\sc i}]-traceable, whereas the hatched region within the CO-bright gas corresponds to ISM that is brighter in [C{\sc i}]~(1-0) than in CO~(1-0) (see Appendix~\ref{app:emission}, including also a comparison with CO~(2-1)). From this simulation, we find that the CO-bright gas covers $\approx10.0\%$ of the molecular gas in area ($\approx8.6\%$ for the CO~(2-1) line), whereas the C{\sc i}-bright covers $\approx38.6\%$. This means that the molecular gas as seen through CO~(1-0) appears to be more clumpy than it truly is, while [C{\sc i}]~(1-0) may cover a wider area of it including the more fluffy molecular gas. We also find that with the adopted detection limit, approximately $95.3\%$ of the total $\rm N(H_2)$ can be measured using the [C{\sc i}]~(1-0) line but only $\simeq73.1\%$ using the CO~(1-0) line ($\simeq70.4\%$ using the CO~(2-1)), illustrating that the [C{\sc i}] line can offer a more accurate estimation of the molecular mass content in Milky Way clouds.

\section{Conclusions}
\label{s:5}

In this paper, we have presented \textsc{raytheia}, a high-performance reverse ray-tracing algorithm designed to efficiently solve three-dimensional direction-dependent equations in astronomical simulations. Our main conclusions are as follow:

\begin{enumerate}
    \item {\sc raytheia} adopts a dual-grid framework in which the native simulation mesh serves as the source grid, while an AMR Cartesian contribution grid is used for ray-walking and contribution accumulation. This decoupled design improves both computational efficiency and methodological flexibility.

    \item The core ray-walking algorithm integrates a leaf-only linear-octree data structure, a DDA traversal method, Morton Code indexing, and the slab method. Together, these features reduce memory overhead, restrict the traversal to the AMR cells actually intersected by a ray, and allow adaptive step sizes that improve ray-walking efficiency in three dimensional simulations.

    \item {\sc raytheia} achieves high-end parallel performance through a hybrid MPI/OpenMP distributed framework and a chunk-to-chunk communication strategy. The strong-scaling result shows a near-ideal speed-up up to 1792 processors, demonstrating that the method is well suited to large-scale, high-resolution simulations.

    \item Across varying angular resolutions, \textsc{raytheia} accurately reproduces the hydrogen column density $N_{\mathrm{H},q}$ distribution of the turbulent cloud, yields minimal deviations in the effective visual extinction $A_{\rm V,eff}$, chemical abundances, and gas temperatures, demonstrating clear convergence as the angular resolution increases. When integrated with \textsc{3d-pdr}, $\mathcal{N}_{\rm rays}=48$ provides a good default compromise between accuracy and computational cost, while even $\mathcal{N}_{\rm rays}=12$ yields chemical abundances and gas temperatures that are highly consistent with higher resolution results, making it a cost-effective alternative for large-scale simulations.

    \item By combining {\sc raytheia} with {\sc 3d-pdr}, we have tested its capabilities in simulating the PDR astrochemistry of a turbulent star-forming cloud. The resulting high-resolution synthetic maps of the upgraded {\sc 3d-pdr} code reveal self-absorption in the [O{\sc i}]~$63\mu$m line within the dense star-forming region, and show that [C{\sc i}] traces a substantially larger fraction of the molecular gas than CO under the adopted detection threshold.
\end{enumerate}

Overall, \textsc{raytheia} provides a powerful, scalable, and computationally efficient tool for tackling direction-dependent problems in Computational Astrophysics, opening new possibilities to more detailed and physically realistic simulations of the ISM.

\section*{Acknowledgements}

The authors thank the anonymous referee for their comments which improved the clarity of the work. The authors thank Richard W\"unsch for the discussions.
This work is supported by the Leading Innovation and
Entrepreneurship Team of Zhejiang Province of China (Grant No. 2023R01008). BALG is supported by the German Research Foundation (DFG) in the form of an Emmy Noether Research Group - DFG project \#542802847 (GA 3170/3-1).

% The Acknowledgements section is not numbered. Here you can thank helpful
% colleagues, acknowledge funding agencies, telescopes and facilities used etc.

%%%%%%%%%%%%%%%%%%%%%%%%%%%%%%%%%%%%%%%%%%%%%%%%%%
\section*{Data Availability}

The data underlying this article will be shared on reasonable request
to the corresponding author.
%The inclusion of a Data Availability Statement is a requirement for articles published in MNRAS. Data Availability Statements provide a standardised format for readers to understand the availability of data underlying the research results described in the article. The statement may refer to original data generated in the course of the study or to third-party data analysed in the article. The statement should describe and provide means of access, where possible, by linking to the data or providing the required accession numbers for the relevant databases or DOIs.



\bibliographystyle{mnras}
\bibliography{refs} % if your bibtex file is called example.bib

%\begin{thebibliography}{99}
%\end{thebibliography}

%%%%%%%%%%%%%%%%%%%%%%%%%%%%%%%%%%%%%%%%%%%%%%%%%%

%%%%%%%%%%%%%%%%% APPENDICES %%%%%%%%%%%%%%%%%%%%%

\appendix

\section{Convergence test on angular resolution}
\label{app:abundances}

\begin{figure*}
    \centering
    \includegraphics[width=0.95\textwidth]{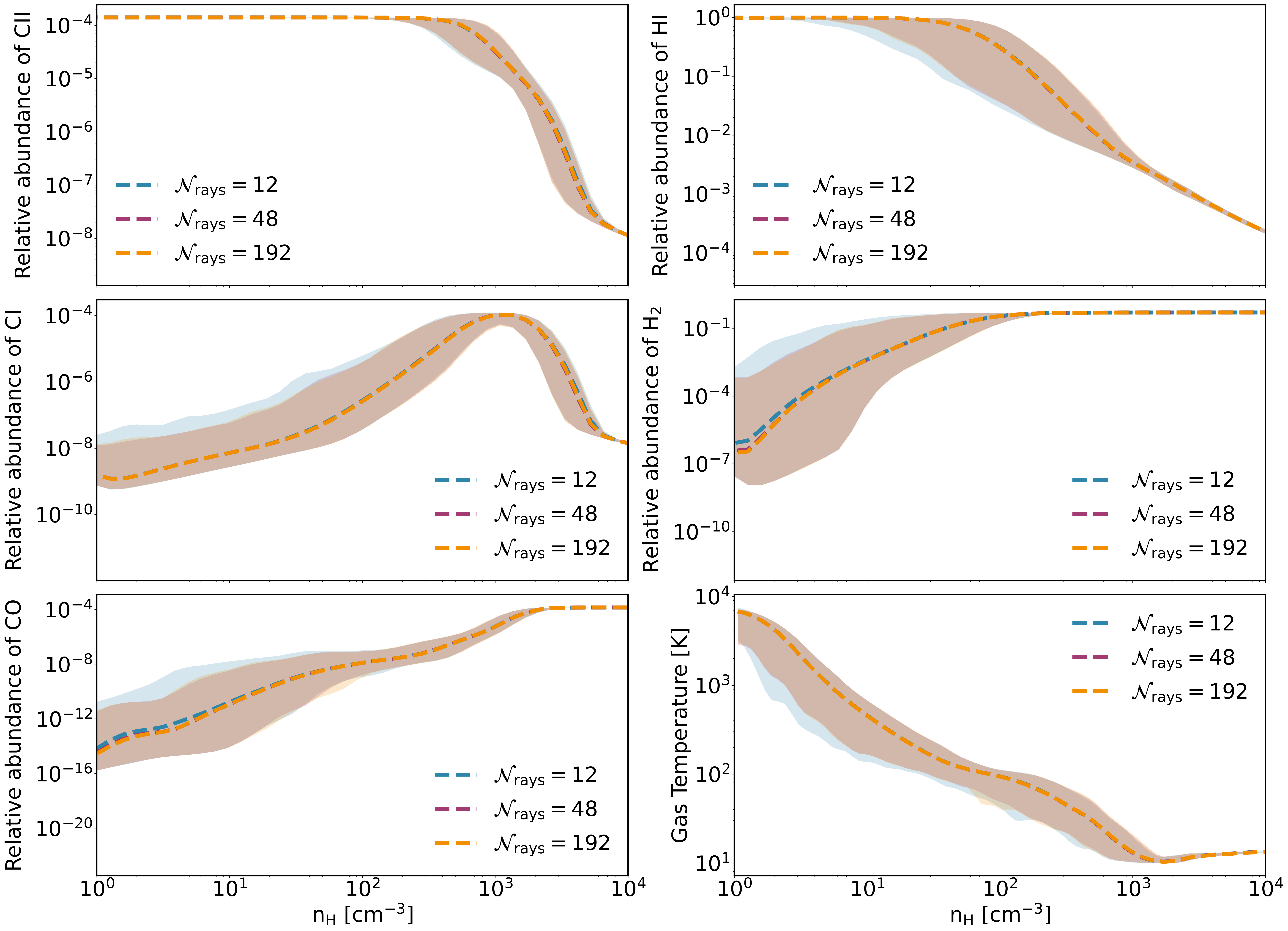}
    \caption{Comparison of the abundances (relative to hydrogen) and gas temperature of the upgraded {\sc 3d-pdr} code with different number of {\sc HEALPix} rays. The shaded area shows the $1\sigma$ standard deviation of abundances and gas temperature when all grid cells are taken into account. The dashed line represents the mean profile of abundances and gas temperature.}
    \label{fig:abundances}
\end{figure*}

Figure~\ref{fig:abundances} presents the chemical and thermal structure of the gas as a function of the total hydrogen number density, $n_{\rm H}$, for different angular resolutions of the radiative transfer, parameterized by the number of {\sc HEALPix} rays, ${\cal N}_{\rm rays}$. The panels show the abundances (relative to hydrogen) of C$^+$, C, CO, H{\sc i}, and H$_2$, together with the gas temperature. For each choice of ${\cal N}_{\rm rays}$, the dashed curves correspond to the mean abundance (or temperature) at a given density, while the shaded regions indicate the full spread of values, reflecting variations in local environmental conditions such as attenuation, geometry, and shielding.

The comparison between different angular resolutions demonstrates excellent convergence of the method. The mean profiles obtained with ${\cal N}_{\rm rays}=48$ and $192$ are nearly indistinguishable over the full density range for all species and for the temperature, indicating that the radiative transfer and shielding are already well captured at moderate angular resolution. Even the lowest resolution case, ${\cal N}_{\rm rays}=12$, reproduces the global trends with remarkable fidelity, showing only modest deviations at the lowest densities, where the radiation field is more anisotropic and therefore more sensitive to angular sampling. The larger spread in the shaded regions at low densities further reflects this increased sensitivity to geometry and line-of-sight variations.

\section{Optical depth effects in CO~(1-0)}
\label{app:tau}

\begin{figure}
    \centering
    \includegraphics[width=0.95\linewidth]{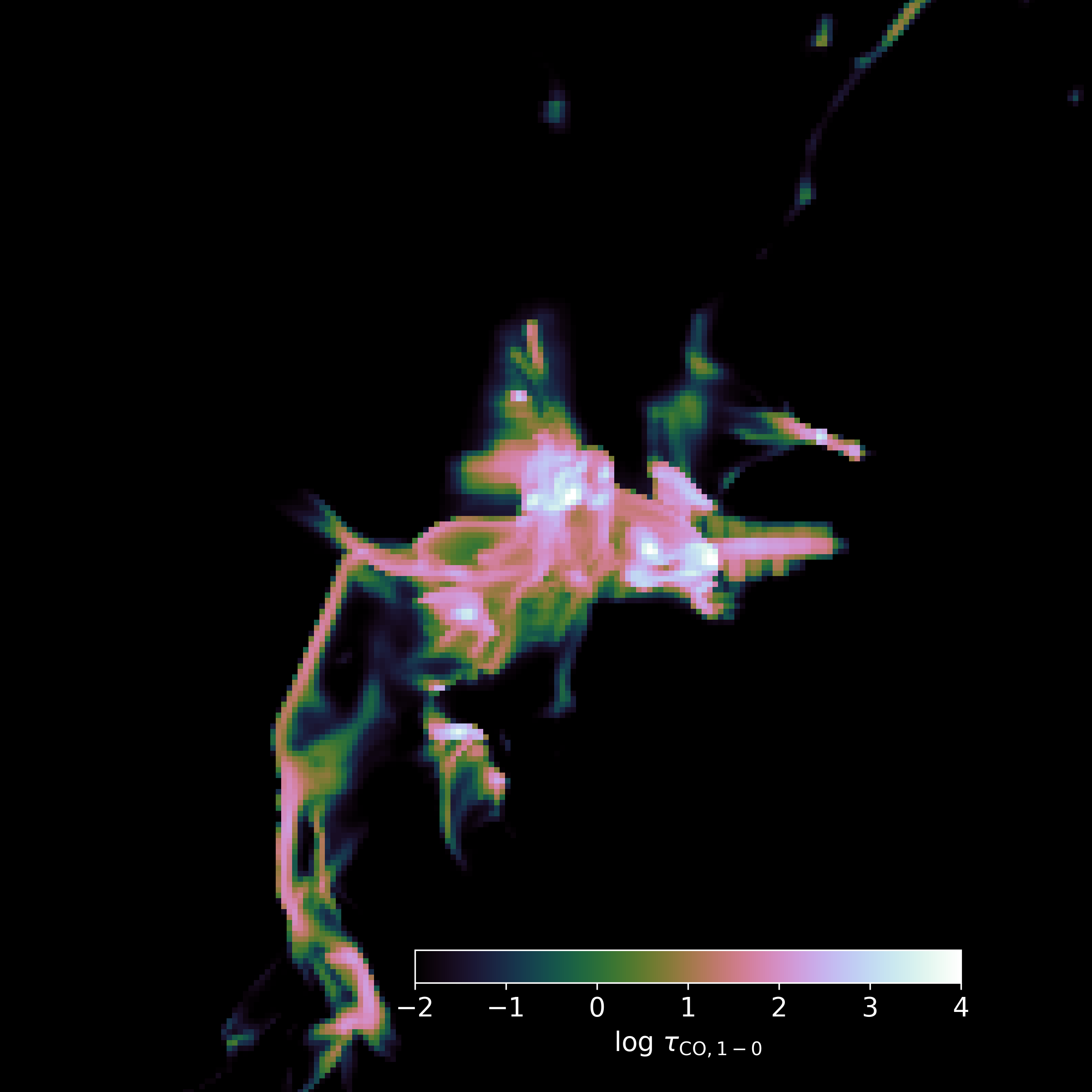}
    \caption{Map of the optical depth of the CO(1–0) transition, $\tau_{\rm CO,1-0}$, in the central high-density region of the simulated cloud. The color scale spans several orders of magnitude, showing that large areas of the collapsing gas reach very high optical depths ($\tau_{\rm CO,1-0}\gtrsim10$), with peak values $\gg10^2$. This widespread optical thickness indicates that the CO(1–0) emission is strongly saturated, with the emergent intensity tracing primarily the $\tau_{\rm CO,1-0}\sim1$ layers rather than the total molecular column density.}
    \label{fig:tau}
\end{figure}

In order to assess the origin of the elevated $X_{\rm CO}$ values obtained in our models, we examine the optical depth of the CO(1–0) transition in the central, high-density region of the cloud. Figure~\ref{fig:tau}  shows a representative map of the line optical depth, $\tau_{\rm CO,1-0}$, across this region. The emission is characterized by widespread and extremely large optical depths, with $\tau_{\rm CO,1-0}\gtrsim10$ over extended areas and peak values reaching $\tau_{\rm CO,1-0}\gg10^2$.

These results indicate that the CO(1–0) line operates in a strongly optically thick regime throughout the dense gas. In this limit, the emergent brightness temperature is no longer sensitive to the total CO column density but instead approaches $T_B \approx J_\nu(T_{\rm ex}) - J_\nu(T_{\rm bg})$, such that the line emission becomes effectively saturated. As a consequence, the velocity-integrated emission, W(CO 1-0) increases only weakly with increasing molecular column density.

This behaviour is directly linked to the structure of the simulated cloud. The bulk of the total gas mass in the model is concentrated within a compact, collapsing region, while most of the remaining volume is occupied by diffuse gas of $n_{\rm H}<1\,\mathrm{cm^{-3}}$. This strong mass concentration leads to a rapid build-up of column density and, consequently, very large optical depths in CO(1–0). In this regime, a significant fraction of the molecular gas becomes effectively hidden behind optically thick layers, and the emergent emission traces primarily the $\tau_{\rm CO,1-0}\sim1$ surfaces in position–velocity space.

The high optical depths demonstrated in Fig.~\ref{fig:tau} therefore provide a natural explanation for the elevated $X_{\rm CO}$ values found in our models. Even though the CO(1–0) transition is close to local thermodynamic equilibrium, the saturation of the line prevents the integrated intensity from scaling linearly with N(H$_2$), leading to systematically higher values of $X_{\rm CO}$ compared to the canonical Milky Way average.

\section{Emission lines vs N(H$_2$)}
\label{app:emission}

\begin{figure}
    \centering
    \includegraphics[width=0.95\linewidth]{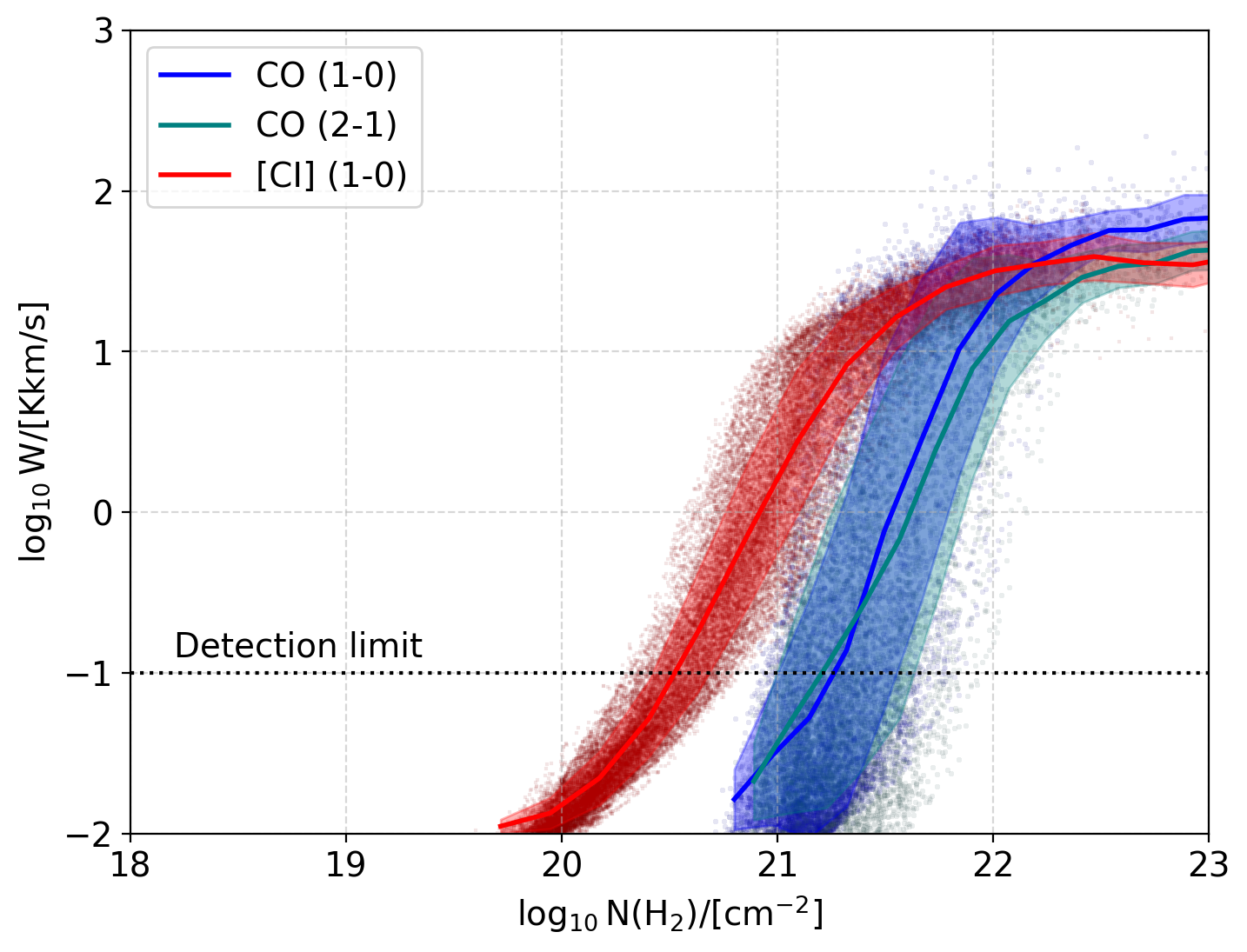}
    \caption{Velocity integrated emission versus the H$_2$ column density for [C{\sc i}]~(1-0) (red colour) CO~(1-0) (blue colour), and CO~(2-1) (teal colour). The solid line corresponds to the mean and the shaded area to $1\sigma$ standard deviation. The scatter points indicate the values of the 2D grid.}
    \label{fig:WH2}
\end{figure}

Figure~\ref{fig:WH2} provides a detailed view of how the velocity–integrated line intensities of CO~(1–0), CO~(2-1) and [C{\sc i}]~(1–0) build up as a function of the molecular hydrogen column density, ${\rm N(H_2)}$. Each point corresponds to an individual line of sight (or pixel) capturing both the intrinsic scatter arising from variations in local physical conditions (e.g. density, temperature, and radiation field) and the systematic trends of the emission with increasing column density.

A clear transition is observed in the range $20 \lesssim \log_{10} {\rm N(H_2)}/{\rm cm^{-2}} \lesssim 21$, where [C{\sc i}]~(1–0) systematically exceeds CO~(1–0) in brightness. This behavior reflects the chemical and radiative structure of PDRs: atomic carbon remains abundant over a broader range of visual extinctions compared to CO, which forms deeper into the cloud where shielding becomes effective. As a result, [C{\sc i}] traces a more extended portion of the molecular gas, including regions where CO is still forming or remains partially photodissociated. In this intermediate column density regime, [C{\sc i}] is a particularly sensitive tracer of H$_2$, especially in environments where CO emission is weak or not yet fully developed.

At higher column densities ($\log_{10} {\rm N(H_2)}/{\rm cm^{-2}} \gtrsim 21.5$), the two tracers converge and eventually CO becomes comparable to or slightly brighter than [C{\sc i}]. This reflects the transition to well-shielded, dense gas where carbon is predominantly locked into CO and the line becomes optically thick, leading to the familiar saturation behavior. The reduced dynamic range of CO in this regime is consistent with its high optical depth, while [C{\sc i}] retains a somewhat broader sensitivity due to its lower opacity. As further shown in Fig.~\ref{fig:WH2}, the CO~(2-1) emission closely follows the behavior of CO~(1-0). This indicates that the $J=2-1$ transition can effectively trace H$_2$-rich gas in dense and well-shielded regions \citep{denBrok23,Teng23}.

The horizontal dashed line indicates the adopted detection limit of $0.1\,{\rm K\,km\,s^{-1}}$. This threshold highlights an important observational implication: a substantial fraction of the [C{\sc i}] emission remains detectable at column densities where CO falls below the sensitivity limit. Consequently, surveys relying solely on CO may miss a non-negligible component of the molecular gas, particularly in diffuse or transitional regions. In contrast, [C{\sc i}] offers a complementary and, in some regimes, superior probe of the H$_2$ distribution. While CO~(1–0) remains a robust tracer of dense, well-shielded molecular gas, [C{\sc i}]~(1–0) provides access to lower column density gas and can serve as a promising alternative tracer of H$_2$-rich gas.

\begin{figure}
    \centering
    \includegraphics[width=0.95\linewidth]{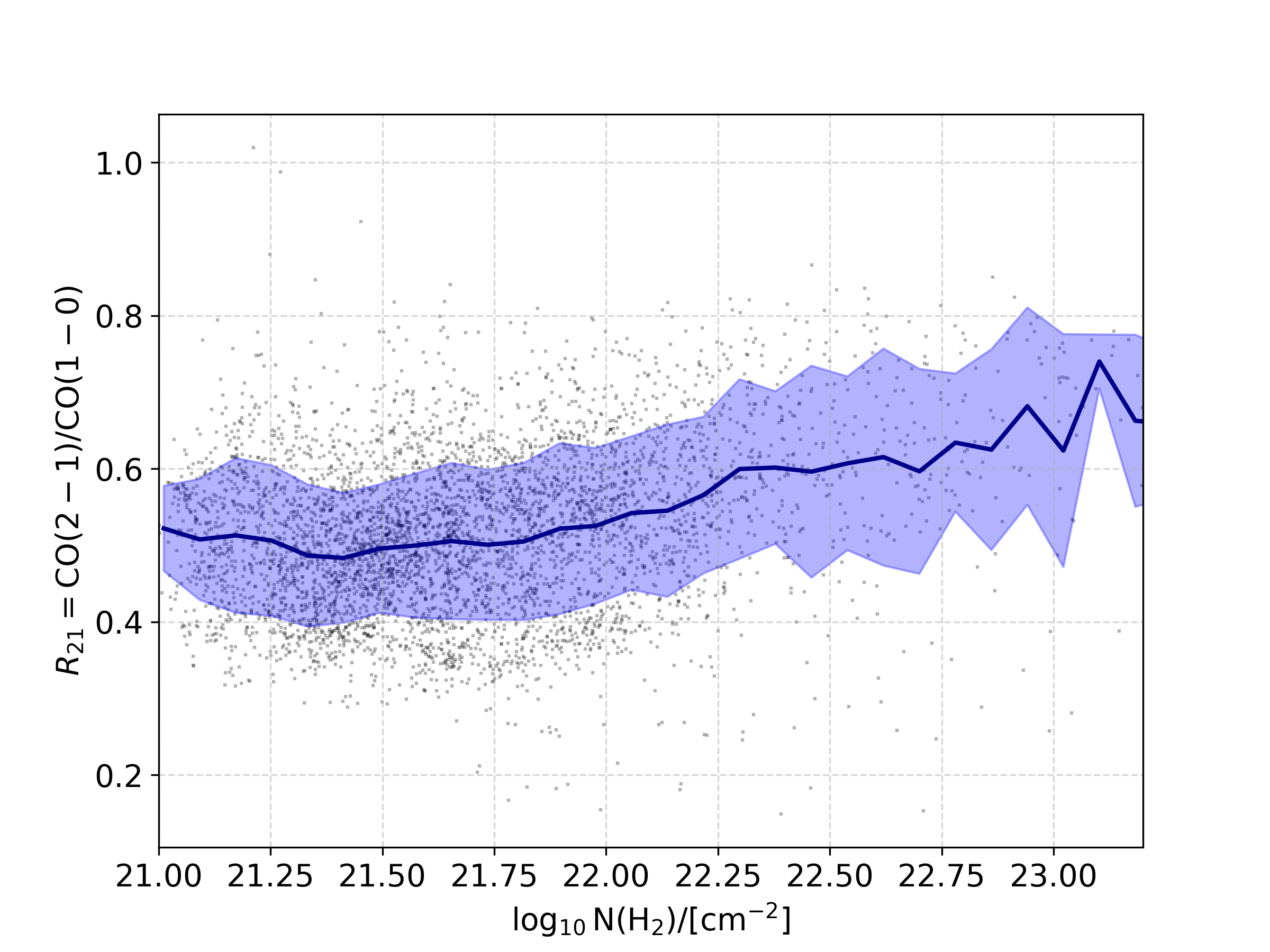}
    \caption{Variation of the CO line ratio $R_{21}$=CO(2-1)/CO(1-0) as a function of the H$_2$ column density. The gradual increase of $R_{21}$ toward higher column densities suggests enhanced excitation in denser and more shielded molecular gas.}
    \label{fig:R21}
\end{figure}

It would also be interesting to explore how the $R_{21}$=CO(2-1)/CO(1-0) emission ratio correlates with N(H$_2$). This is shown in Fig.~\ref{fig:R21} where the detection limit of $0.1\,{\rm K}\,{\rm km/s}$ is considered for both these lines. We find that $R_{21}$ spans values between $\sim0.4-0.8$ with mean values of $\sim0.5-0.7$, consistent with observations of nearby molecular clouds and star-forming galaxies \citep[e.g.][]{Penaloza18,denBrok21,Teng23}. Our model predicts a gradual increase of $R_{21}$ with increasing N(H$_2$), suggesting enhanced excitation toward denser and more shielded molecular gas. This behavior is expected, since the CO~(2-1) transition preferentially traces warmer and denser gas compared to CO~(1-0), leading to brighter CO~(2-1) emission in star-forming environments. The fact that the ratio remains mostly below unity indicates that the gas is not fully thermalized.

The scatter representing individual cells is also present for a fixed N(H$_2$), indicating that the excitation conditions cannot be described by molecular column density alone. Variations in local gas density, kinetic temperature and velocity structure all contribute to the spread in $R_{21}$. Such scatter is commonly observed in resolved molecular cloud studies \citep[e.g.][]{Koda25,Lee2026}.

% If you want to present additional material which would interrupt the flow of the main paper,
% it can be placed in an Appendix which appears after the list of references.

%%%%%%%%%%%%%%%%%%%%%%%%%%%%%%%%%%%%%%%%%%%%%%%%%%

% Don't change these lines
\bsp	% typesetting comment
\label{lastpage}
\end{document}